\documentclass[preprint,12pt]{elsarticle}

\usepackage{amssymb}
\usepackage{amsmath}

\usepackage{lineno}

\newcounter{bla}

\usepackage{natbib}

\journal{Computer Physics Communications}

\begin{document}

\begin{frontmatter}



\title{PyFrac: A planar 3D hydraulic fracture simulator}


\author[a]{Haseeb Zia}
\author[a]{Brice Lecampion} 

\address[a]{Ecole Polytechnique F\'ed\'erale de Lausanne, Geo-Energy Lab, Gaznat chair on Geo-Energy, School of Architecture, Civil \& Environmental Engineering, EPFL-ENAC-IIC-GEL, Station 18, Lausanne, CH-1015, Switzerland}

\begin{abstract}
Fluid driven fractures propagate in the upper earth crust either naturally or in response to engineered fluid injections. The quantitative prediction of their evolution is critical in order to better understand their dynamics as well as to optimize their creation. 
We present an open-source Python implementation of a hydraulic fracture growth simulator based on the implicit level set algorithm originally developed by Peirce \& Detournay  (2008) -- "An implicit level set method for modeling hydraulically driven fractures". {\it Comp. Meth. Appl. Mech. Engng}, {\bf197}(33-40):2858–-2885. This algorithm couples a finite discretization of the fracture with the 
use of the near tip asymptotic solutions of a steadily propagating semi-infinite hydraulic fracture. This allows to resolve the multi-scale processes governing hydraulic fracture propagation accurately, even on relatively coarse meshes. 
We present an overview of the mathematical formulation, the numerical scheme and the details of our implementation. A series of problems including a radial hydraulic fracture verification test, the propagation of a height contained hydraulic fracture, the lateral spreading of a  magmatic dyke and an example of fracture closure are presented to demonstrate the capabilities, accuracy and robustness of the implemented algorithm.
\end{abstract}


\end{frontmatter}


\noindent {\bf Program Summary} \\
\begin{small}
\noindent {\em Program Title: PyFrac}                                          \\
\noindent {\em Licensing provisions: GPLv3}                                   \\
\noindent {\em Programming language: Python}                                   \\
\noindent {\em Nature of problem:} 
Simulation of the propagation and closure of a planar three-dimensional hydraulic fracture driven by the injection of a Newtonian fluid in a material having heterogeneous fracture toughness under a non-uniform in-situ stress field. 

\noindent {\em Solution method:}  
The fully coupled hydro-mechanical moving boundary problem is solved combining a finite volume scheme for lubrication flow with a boundary element method for elasticity. The algorithm couples  a finite scale discretization of the fracture with the near-tip asymptotic solution of a steadily moving hydraulic fracture.  The fracture front is tracked via a level set approach using a fast marching method.


\end{small}

\vspace{0.75cm}

\section{Introduction}

Hydraulic fractures (HFs) are a class of tensile fracture propagating in rocks under pre-existing compressive stress in response to the injection or release of pressurized fluid \cite{Deto16}. 
They are routinely engineered in order to increase the production of oil and gas wells \cite{EcNo00}. Hydraulic fractures are also used in the pre-conditioning of ore body mined via block caving techniques \cite{VaJe02,HeSu2016}. Compensation grouting is another example of their application in civil engineering \cite{EsDr00}. HFs also occur naturally as dykes propagating from deep pressurized magma chamber \cite{RiTa15} or as fracture propagating at glacier beds following sudden fluid discharge \cite{Van07,TsRi10}. 
Quantitative estimate of the dynamics and extent of hydraulic fractures is critical in practical applications in order to optimize the engineering design. This is typically done with the help of numerical models. 
In addition, numerical modelling can also help in understanding  hydraulic fracture growth in non-trivial configurations.

Numerical modelling of hydraulic fractures has been an active area of research since the end of the 1950s. The mathematical models have evolved from simple geometries with ad-hoc growth physics to sophisticated three dimensional models -  see \cite{AdSi07,LeBu18} for the most recent reviews. 
 The numerical modeling of the propagation of hydraulic fracture is extremely challenging due to a number of reasons. In addition to the intrinsic moving boundary nature of the problem, the coupling between the lubrication fluid flow inside the fracture and the elastic deformation of the rock (non-local by essence) is extremely non-linear as the fracture hydraulic transmissivity increases with the cube of the local fracture width. 
 Such a hydro-mechanical coupling yields a complex multiscale structure of the solution in the near tip region where the classical linear elastic fracture mechanics asymptote can reduce to a small boundary layer near the tip while a viscous asymptote control the far-field behavior. An intermediate asymptote due to fluid leaking off in the surrounding rock can also appear - see \cite{GaDe11,BuDe08} for detailed solutions and experimental validation. This multi-scale structure of the solution near the propagating hydraulic fracture front is known to control the propagation of finite hydraulic fractures which exhibit a competition between the dissipative processes associated with fluid flow and fracture creation as well as between the amount of fluid leaking off the fracture compared to the amount stored within the fracture \cite{Gara09}.   
 Numerical models of HF growth must therefore properly resolve these different lengthscales near the fracture front in order to yield accurate results. This is particularly challenging numerically as the extent  of the different asymptotic regions can vary widely as function of the rock and injection properties - therefore requiring extremely fine meshes in some cases. 

We present a Python implementation of a particularly efficient numerical scheme for hydraulic fracture (HF) propagation denoted as the implicit level set algorithm (ILSA) \cite{PeDe08,Peir15,Peir16,DoPe17}. 
The scheme elegantly couples the near-tip asymptotic solution of a steadily moving hydraulic fracture \citep{GaDe11} (valid in the near-tip region) with a finite discretization of the fracture.  By using the near-tip HF asymptotic solution,  the challenging numerical resolution of the multiscales structure of the solution near the tip is avoided altogether.
As a result, this allows to obtain highly accurate solutions even on relatively coarse meshes as compared to other fracture propagation algorithms (see \cite{LePe13,LeBu18} for comparisons).

Our Python  implementation also includes a number of extensions to the original ILSA scheme. In particular, the developed solver
includes 1) the capability to advance the fracture front implicitly, explicitly or in a predictor-corrector fashion \cite{ZiLe19ExIm}, 2) the modification of the lubrication flow to take into account the possible transition to turbulent flow \cite{LeZi19}, 3) the possibility to account for an anisotropy of fracture toughness and elasticity \cite{ZiLe18,MoLe19}, and finally 4) the capability to handle closure of the fracture after the end of pumping. 

In the following, we briefly describe the underlying mathematical model of HF growth, its solution in the context of the implicit level set algorithm and discuss some details of our implementation. Several examples are then discussed in order to illustrate the accuracy and capabilities of the developed numerical code. 

\section{Mathematical model}

PyFrac solves the equations of the classical linear elastic hydraulic fracture problem for a three-dimensional planar fracture. We briefly recall below the governing equations for such class of problems and refer to 
\cite{Deto16,LeBu18} for a more detailed description of the underlying physical assumptions.

\subsection{Elastic deformation}

For a pure opening mode planar fracture (mode I), the quasi-static balance of momentum of the medium reduces to a single  hyper singular boundary integral equation relating the fracture width  $w$ (normal displacement discontinuity) and the normal component of the traction vector. In the case of a planar fracture of area $A(t)$ (evolving with time) in a homogeneous isotropic material, it further reduces to
(see \cite{CrSt83,HiKo96} for details)
\begin{equation}
T(x,y,t)-\sigma_{o}(x,y)=-\frac{E'}{8\pi}\int_{A(t)}\frac{w(x',y',t) \text{ d}A(x',y')}{[(x'-x)^{2}+(y'-y)^{2}]^{3/2}}.\label{eq:Elasticity}
\end{equation}
where $T$ and $\sigma_o$ are the normal components of the applied traction and the far-field in-situ compressive stress respectively. 
We  account for the fact that the fracture opening $w$ can not be negative. More precisely, upon fracture creation, the fracture may close but exhibit a residual aperture $w_a$ taken as the minimum between the maximum opening encountered thus far at this position and a value related to the intrinsic roughness of the created fracture $w_r$:\\ $w_a=\textrm{min}\left(\textrm{max}(w), w_r \right)$. This results in the following contact conditions 
\begin{equation}
    (w -w_a)\ge 0 \qquad (T-p) (w-w_a) =0 
    \label{eq:width_constraint}
\end{equation}
which states that if the fracture is mechanically open at a given location, the corresponding normal traction on the fracture faces $T(x,y,t)$ equals the fluid pressure $p(x,y,t)$. 

\subsection{Lubrication flow inside the fracture}

The fluid flow inside the fracture obeys the lubrication approximation \cite{Batc67}. The width averaged mass conservation for a slightly compressible liquid reduces to (see e.g.~\cite{LeBu18})
\begin{equation}
\frac{\partial w}{\partial t}+c_{f}w\frac{\partial p}{\partial t}+\nabla\cdot\mathbf{q}+v_{L}=Q(x,y)\delta(x,y),\label{eq:lubrication}
\end{equation}
where $v_L$ denotes the velocity of the fluid leaking out of the two opposite faces of the fracture, and $\mathbf{q}$ is the fluid flux within the fracture. 
Similarly, for such a lubrication flow, the width averaged balance of momentum of the fluid reduces to Poiseuille's law. Accounting for the possible appearance of turbulent flow (but still neglecting inertial terms), the fluid flux $\mathbf{q}=w\times \mathbf{v}$ is directly related to the fluid pressure gradient as \cite{ZiLe17,DoPe17}:
\begin{equation}
\mathbf{q}=\frac{-w^{3}}{12\mu\,\tilde{f}\left(Re_{Deq},w_{R}/w\right)}(\nabla p+\rho\mathbf{g}),\label{eq:flux}
\end{equation}
where the reduced Fanning friction factor $\tilde{f}$ is defined as:
\begin{equation}
\tilde{f}\left(Re_{Deq},\frac{w_{R}}{w}\right)=f\left(\frac{4}{3}Re,\frac{w_{R}}{w}\right)/f_{laminar}.\label{eq:friction_factor}
\end{equation}
It captures the possible transition to turbulent flow inside the fracture as function of the local Reynolds number $Re=\rho w v /\mu $ and the roughness length scale $w_R$: $f$ is the Fanning friction factor expression for turbulent flow in pipe (function of the Reynolds number) and $f_{laminar}=64/Re$ is the laminar expression such that $\tilde{f}=1$ for laminar flow. Different models for Fanning friction are available in our implementation - notably the one described in \cite{YaDo10,YaJo09}.

The leak off fluid velocity $v_{L}$ is evaluated using the Carter's leak off model (see e.g.~\cite{LeBu18} for discussion):
\begin{equation}
v_{L}=\frac{2 C_{L}(x,y)}{\sqrt{t-t_{0}(x,y)}}.\label{eq:er_leakoff}
\end{equation}
where $C_L$ [$L.T^{-1/2}$] is the Carter's leak-off coefficient which depends on the rock and fracturing fluid properties.

\subsection{Boundary Conditions\label{subsec:Boundary-Conditions}}

PyFrac assumes that the fluid and fracture front coincides: a condition typically encountered when the 
in-situ normal compressive stress $\sigma_o$ is sufficiently large (see \cite{GaDe00} for discussion). As a result, at the fracture front, beside the condition of zero fracture width, the component of the fluid flux
$\mathbf{q}(x,t) $ normal to the fluid front also vanishes:
\begin{equation}
w(x_{c},t)=0,\;\;\;\mathbf{q}(x_{c},t)\cdot \mathbf{n}(x_{c},t)=0,\;\;\;x_{c}\in C(t),\label{eq:boundary_conditions}
\end{equation}
where $C(t)$ denotes the fracture front at time $t$ and $\mathbf{n}(x_{c},t) $ its corresponding normal.
 Moreover, the hydraulic fracture is assumed to be propagating in quasi-static equilibrium.
As a result, the stress intensity factor everywhere along the fracture front is equal  to (or below for a stagnant front) the fracture toughness of the rock. This results in the following propagation condition:
\begin{eqnarray}
(K_{I}(x_{c},t)-K_{Ic}(x_{c},\alpha)) \le 0 \\
(K_{I}(x_{c},t)-K_{Ic}(x_{c},\alpha)) \times V(x_c) =0\;\;\;x_{c}\in C(t).\label{eq:propagation_condition}
\end{eqnarray}
where $V(x_c)\ge 0 $ is the local fracture propagation velocity.
The fracture toughness of the material $K_{Ic}$ can possibly be function of position (inhomogeneous material) as well as of the propagation direction $\alpha$ in the case of a material with an anisotropic fracture toughness (see \cite{ZiLe18} for details).

\section{Numerical Solution} 
\begin{figure}
\begin{centering}
\includegraphics[scale=1.1]{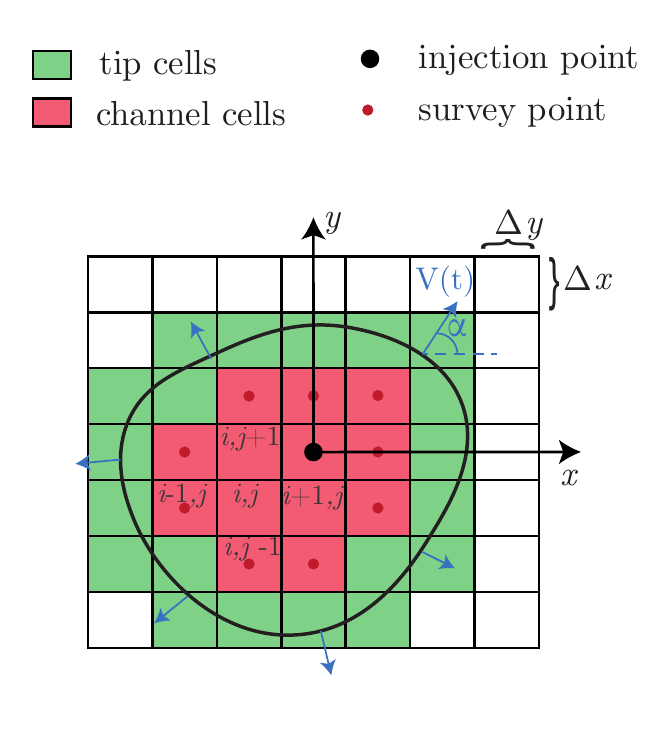}
\par\end{centering}
\caption{Schematic of the finite discretization of the fracture plane  with the fracture front cutting through the background Cartesian grid.  At any time, the cells are classified as either tip (near the front) or channel cells. Among the channel cells, the center of the cells adjacent to the tip cells are taken as survey points and are used to couple the finite discretization with the near-tip hydraulic asymptotic solution.
}
\label{fig:cells_classification}
\end{figure}

In this section, We outline some details of our implementation of ILSA. We refer to the description given in \cite{PeDe08,Peir15,DoPe17} for more details.

\subsection{Discretization}

The hydraulic fracture is discretized using a fixed Cartesian mesh with rectangular cells of sizes $\Delta x$, $\Delta y$ - see figure \ref{fig:cells_classification}. The algorithm marches forward in time from a known solution at time $t^n$ which consists of the location of the fracture front (intersecting the background grid), width and fluid pressures at the center of the cells located inside the fracture. 

Using the distributed dislocation technique, the elasticity equation (\ref{eq:Elasticity}) is collocated at the center of each cell within the current fracture footprint assuming a piece-wise constant value of fracture width in each cell. It results in a  dense linear system for an open fracture loaded by a fluid (where $T=p$ in equation (\ref{eq:Elasticity})). The fluid 
pressure $p_{i,j}$ at cell $(i,j)$ located in an open part of the fracture (see figure \ref{fig:cells_classification}) is linearly related to the opening in all the other cells 
\begin{equation}
p_{i,j}-\sigma_{o\,i,j}=\mathbb{E}_{i,j;k,l}w_{k,l},  \label{eq:ElastOperator}
\end{equation}
where $\mathbb{E}_{i,j;k,l}$ is the elastic contribution of cell $(k,l)$ on cell $(i,j)$ and summation is performed on repeated indices. 
If the fracture is mechanically closed, the width $w_{i,j}$ equals the residual aperture $w_a$, and the corresponding traction $T_{i,j}$ is now unknown (see eq.~(\ref{eq:width_constraint})).

The lubrication equation (\ref{eq:lubrication}) 
is discretized using a cell centered finite volume method. Using a backward-Euler time integration scheme, one obtains the following equation for cell $(i,j)$ over the time step of size $\Delta t$: 
\begin{equation}
\Delta w_{i,j}=[A\,p]_{i,j}-[C\Delta p]_{i,j}+G_{i,j}+\Delta t\,Q_{i,j}-\mathcal{L}_{i,j},\label{eq:LubrOper}
\end{equation}
where the fluid flux across the cell edges is approximated by central finite difference resulting in a five point stencil:
\begin{eqnarray}
[A\,p]_{i,j} &=&
\frac{\Delta t}{\Delta x^{2}}\left(K_{i+1/2,j}p_{i+1,j}-(K_{i+1/2,j}+K_{i-1/2,j})p_{i,j}+K_{i-1/2,j}p_{i-1,j}\right) \nonumber \\
&+&\frac{\Delta t}{\Delta y^{2}}\left(K_{i,j+1/2}p_{i,j+1}-(K_{i,j+1/2}+K_{i,j-1/2})p_{i,j}+K_{i,j-1/2}p_{i,j-1}\right)
\end{eqnarray}
The fracture fluid transmissivity $K_{i-1/2,j}$ at the cell edge $(i-1/2,j)$ (and similarly for the other edges) is given by 
\begin{equation}
K_{i-1/2,j}=\frac{w_{i-1/2,j}^{3}}{12\mu\,\tilde{f}(Re_{Deq\,i-1/2,j},w_{R}/w_{i-1/2,j})},
\end{equation}
where the width and Reynolds number are averages of the two neighboring cells. These transmissivities are non-linearly dependent on the current estimate of fracture width.

For a gravity vector aligned along the $y$ axis of the grid, the gravity term $G_{i,j}$ is given by 
\begin{equation}
G_{i,j}=\frac{\Delta t}{\Delta y}(K_{i,j+1/2}-K_{i,j-1/2})\rho g,
\end{equation}
while the effect of fluid compressibility is strictly local and reads
\begin{equation}
[C\Delta p]_{i,j}=c_{f}\left(w^n_{i,j}+\frac{\Delta w_{i,j}}{2}\right) \Delta p_{i,j}.
\end{equation}
 $Q_{i,j}$ contains the fluid injection rate (only non-zero in the injection cell).
 The leak-off contribution $ \mathcal{L}_{i,j}$ for cell $(i,j)$ over the time-step is approximated as \cite{PeDe08}:
 \begin{equation}
     \mathcal{L}_{i,j} = 4 C_L \Delta t \left( \sqrt{t^n+\Delta t-t_{o\,i,j}} - \sqrt{t^n-t_{o\, i,j}} \right)
 \end{equation}
where $t_{o\,i,j}$ is the time at which the fracture front has first passed through the center of cell $(i,j)$ (also refereed to as the trigger time for leak-off).


\subsection{Elasto-hydrodynamics solver}

For a {\it known trial} position of the fracture front at time $t+\Delta t$, the non-linear 
 coupling between the discretized lubrication (\ref{eq:LubrOper}) and elastic (\ref{eq:ElastOperator}) equations  can be re-written in a matrix form. Taking the increment of 
width and pressure in all cells within the fracture footprint as the primary unknowns, one obtains the following non-linear system when no width constraints are active:
\begin{equation}
\left[\begin{array}{cc}
\mathbb{E} & -\mathbf{I}\\
\mathbf{I} & -\mathbb{L}(\mathbf{\Delta  w}) 
\end{array}\right]\left[\begin{array}{c}
\mathbf{\Delta w}\\
\mathbf{\Delta p}
\end{array}\right]=\left[\begin{array}{c}
\mathbf{0}\\
\underbrace{\mathbf{A}(\mathbf{\Delta w})\cdot \mathbf{p}^n+\mathbf{G}(\mathbf{\Delta w})+\mathbf{S}}_{\mathbf{F_L}}
\end{array}\right].
\label{eq:non_linear_system}
\end{equation}
In the previous equation, $\mathbf{I}$ denotes the identity matrix.
The elasticity block $\mathbb{E}$ is dense, while $\mathbb{L}(\mathbf{\Delta w})=\mathbf{A}(\mathbf{\Delta w})-\mathbf{C}(\mathbf{\Delta w})$ is sparse (notably the effect of compressibility $\mathbf{C}$ is strictly diagonal). We have also highlighted the non-linear dependence of $\mathbb{L}$ on the current fracture width increment, and defined $\mathcal{S}=\Delta t\,\mathbf{Q}-\mathcal{L}$ combining the injection sources and leak-off sink terms.
 
As previously mentioned, the implicit level set  algorithm incorporates the near tip asymptotic solution for a steadily moving hydraulic fracture near the fracture front.
This is done by identifying the  cells intersecting with the fracture front (denoted as tip cells)
and the cells within the fracture apart from the tip cells (called channel cells). 
The fracture widths of the tip cells are imposed according to the  HF tip solution which depends on the current estimate of the local fracture velocity.  
The pressure in the  flow equation (\ref{eq:LubrOper}) is substituted with width using
the elasticity equation (\ref{eq:ElastOperator}) for mechanically open channel cells (denoted with a superscript $C$). 
In addition to imposing the fracture width in the tip
cells, we also enforce the minimum width constraint  (\ref{eq:width_constraint}) everywhere - and denote the corresponding set of cells with active constraints with a superscript $\mathcal{A}$. 

After imposing the width according to the HF tip asymptote and the active minimum width constraints in the set of tip (denoted with a superscript $T$) and active cells (superscript $\mathcal{A}$)  respectively,  
the nonlinear system (\ref{eq:non_linear_system}) can be re-written to solve for the increment of width $\Delta w$ in the channel cells  and increment of fluid pressure $\Delta p$ in the tip cells ($T$)  and the cells ($\mathcal{A}$) with an active width constraint. The final non-linear system can be expressed in the following format highlighting the different sub-blocks:
\begin{eqnarray}
 \begin{bmatrix}
\mathbf{I}^{CC}-\mathbb{L}^{CC} \mathbb{E}^{CC} & -\mathbb{L}^{CT} & -\mathbb{L}^{C\mathcal{A}}\\
-\mathbb{L}^{TC}\mathbb{E}^{CC} & -\mathbb{L}^{TT} & -\mathbb{L}^{T\mathcal{A}}\\
-\mathbb{L}^{\mathcal{A}C}\mathbb{E}^{CC} & -\mathbb{L}^{\mathcal{A}T} & -\mathbb{L}^{\mathcal{AA}}
\end{bmatrix} 
\begin{bmatrix}
 \mathbf{\Delta w}^C \\
 \mathbf{\Delta p}^T  \\
 \mathbf{\Delta p}^{\mathcal{A}}
\end{bmatrix} 
&=&
\begin{bmatrix}
 \mathbf{F_L}^C +\mathbb{L}^{CC} \mathbf{b}^C \\
 \mathbf{F_L}^T -\mathbf{\Delta w}^T +\mathbb{L}^{TC} \mathbf{b}^C\\
 \mathbf{F_L}^{\mathcal{A}}-\mathbf{\Delta w}^\mathcal{A}+\mathbb{L}^{\mathcal{A}C} \mathbf{b}^C
\end{bmatrix} 
\label{eq:elastoHydrodynamic}
\end{eqnarray}{}
where the different matrix sub-blocks are defined with respect to the channel ($C$), tip ($T$) and active ($\mathcal{A}$) cells. We have also defined 
\begin{equation}
    \mathbf{b}^C = \mathbb{E}^{CT} \mathbf{\Delta w}^T +  \mathbb{E}^{C\mathcal{A}} 
    \mathbf{\Delta w}^\mathcal{A}
\end{equation}
and the increment of width in the tip and active cells are simply given by:
\begin{eqnarray}
 \mathbf{\Delta w}^T &=& \mathbf{w}^T - \mathbf{w^n}^T \\
 \mathbf{\Delta w}^\mathcal{A} &=& \mathbf{w_a}^\mathcal{A} - \mathbf{w^n}^\mathcal{A}, 
 \end{eqnarray}
where $\mathbf{w}^{T}$ represents the vector of
width in the tip cells evaluated using the HF tip asymptote \cite{GaDe11} and $\mathbf{w_a}^{\mathcal{A}}$
is the vector of minimum residual width. 
The vectors (e.g. $\mathbf{F_L}^C$) on the right hand side  of eq.~(\ref{eq:elastoHydrodynamic}) are short notation for the right hand side appearing in the system of eq.~(\ref{eq:non_linear_system}).

The non-linear system (\ref{eq:elastoHydrodynamic}) is solved iteratively using a simple fixed-point scheme which has proven to be robust and accurate. Convergence is reached when the L2 norm of subsequent 
estimates of the increment of width and pressure are below a prescribed tolerance (in relative term) - typically $10^{-6}$. The inequality constraints are checked after convergence and the set of active cells updated if required (and the system subsequently re-solved until convergence of the active set).
Due to its non-linear nature and the fact that the previous system needs to be solved for each trial position of the fracture front, it is the most critical part of the solver from a computational point of view.

It is also worthwhile to note that in the case of an inviscid fluid (zero viscosity / toughness dominated propagation), 
the fluid pressure is uniform inside the fracture (in the absence of gravity). In that limiting case, a simpler set of equations can be solved combining elastic deformation and global volume balance in order to solve for increment of width and a single fluid pressure increment (see e.g.~\cite{PeDe08} for details). 

\subsection{The fracture propagation algorithm}

The fracture front is represented by a level set function and its new position at the end of the time step is obtained iteratively in a fully implicit manner in the original ILSA scheme \cite{PeDe08}.

Once the non-linear elasto-hydrodynamics system (\ref{eq:elastoHydrodynamic}) has been solved for a given trial position of the fracture front, the estimate of the new width in the cell just behind the tip cells (survey points in figure \ref{fig:cells_classification}) are used in combination with the HF tip solution in order to obtain the local closest distance $s$ from the survey point to the fracture front. This is performed by inverting the HF tip asymptotic solution giving the fracture width as function of the closest distance to the fracture tip. The closest distance to the fracture front obtained in all the cells behind the tip cells provide an initial condition to solve for the signed distance to the fracture front (i.e. the level set function) in all the grid cells. The solution of this Eikonal equation is performed via a fast marching method. The fracture front can then be reconstructed using a piece-wise linear approximation within each cell. Subsequently, the width in the tip cells for this new position of the fracture front can be imposed using the HF near tip asymptotic solution (using the local fracture front velocity). More precisely, the volume of the tip cells are prescribed to ensure proper volume conservation. The algorithm then re-solve the non-linear elasto-hydrodynamics system to obtain a new estimate of the fracture width increment and tip pressure. 
Convergence is reached when subsequent estimate of the level set function at all survey points falls below a given tolerance (in relative term) - typically $10^{-3}$.

It is interesting to point out that beside a fully implicit scheme, we also provide an explicit as well as predictor-corrector version of the scheme. In the original fully implicit version of the scheme, 
the first trial position of the new fracture front is kept as its value at the end of the previous time-step. The fully explicit version estimate the new position of the fracture front from the local velocities obtained 
at the end of the previous time-step (and thus no iteration on the fracture front position are performed). The predictor-corrector version subsequently iterate from the trial position obtained explicitly. More details and comparisons of the difference scheme are discussed in \cite{ZiLe19ExIm}. By default, PyFrac uses a predictor-corrector scheme but such a choice can be modified by the user if desired. 

\begin{figure}
   \makebox[\textwidth]{
     \includegraphics[scale=0.8]{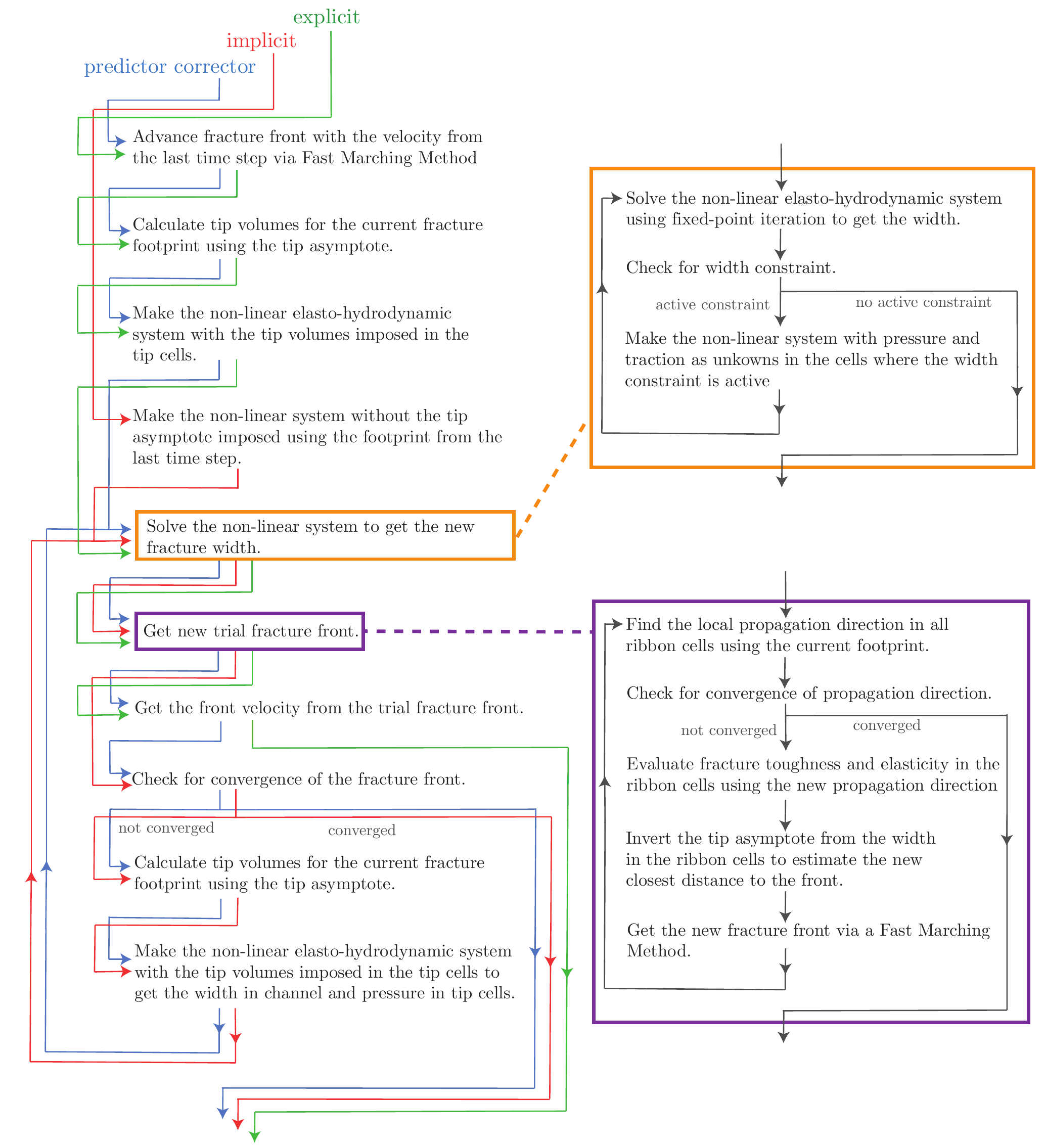}
   }
   \caption{The algorithm used by PyFrac to advance a time step. The predictor
corrector, implicit and explicit front advancing schemes are shown
in blue, red and green colors respectively.\label{fig:flow_chart}}
\end{figure}

A complete summary of the algorithm over one time-step is shown in the form of a flow chart in figure \ref{fig:flow_chart}. 
Note that the value of the time-step is automatically adjusted from the current knowledge of the fracture front velocity - see \cite{ZiLe19ExIm} for more details.

\subsection{Fracture closure}

Fracture closure is solely modeled via the contact condition (\ref{eq:width_constraint}) at the level of the grid.
The algorithm classifies the fracture front as either propagating or stagnant. 
In other words, the fracture front described by the level set does not recede but the fracture can close. 
The fracture is assumed closed in the cells where the faces of the fracture come into contact: when the fracture width becomes equal to the minimum residual width $w_r$ - a given input akin to a material property. Although the fracture front does not recede, there is a front of closing cells which can be seen as a receding front. The direction of this ‘receding front’ is resulting from the coupling between fluid flow / leak-off / elasticity and the contact condition via the injection flow rate history. 
 
 For a stagnant fracture front, the fracture propagation condition is not fulfilled anymore: the local stress intensity factor $K_I$ is below the fracture toughness $K_I<K_{Ic}$. Following a procedure discussed in \cite{DoPe17}, the stress intensity factor is then computed from the width of the ribbon cells. The width of the corresponding tip cells are set according to the linear elastic fracture mechanics asymptote using the computed $K_I$: $w=\sqrt{32/\pi}\frac{K_I}{E^\prime} s^{1/2}$, with $s$  the distance normal to the fracture front.  The tip element volumes, to be imposed in the tip cells, are computed by integrating this width in a similar way than for a propagating front (see \cite{DoPe17} for details).


\section{Implementation}

PyFrac makes extensive use of NumPy \cite{numpy} and SciPy \cite{scipy} routines. The implementation
details of the computationally extensive routines of PyFrac are briefly
discussed below.

\paragraph{Assembly of the Elasto-hydrodynamic system}

The elasto-hydrodynamic system (\ref{eq:elastoHydrodynamic}) requires multiple matrix products of the dense matrix resulting from the boundary integral elastic equation with the sparse finite lubrication matrix (resulting from the  five point stencil finite difference), which is function of the current width estimate.
 PyFrac uses the compressed sparse column matrix provided
by SciPy for the lubrication matrix and the standard
2-dimensional NumPy array for the dense elastic  matrix. The dot product routine provided by SciPy for sparse matrix product is used for efficiency.

\paragraph{Solution of the Elasto-hydrodynamic system}

The non-linear elasto-hydrodynamic system is solved via fixed
point iterations, which converts it into a series of linear systems. PyFrac uses the linear solver provided by NumPy, which is basically a Python wrapper
for the highly efficient direct linear solver provided by LAPACK.

\paragraph{Root finding}

The HF tip asymptotic solution is evaluated by an efficient approximation provided by \citet{DoPe15}, given in the form of an implicit function. To evaluate the tip asymptote as well as to invert it, root finding is required for both inverting the tip asymptote and to evaluate its integral over the tip cell. PyFrac uses the implementation of Brent's method provided by SciPy to find the root of the implicit function.

\subsection{Memory requirements}

PyFrac is a memory intensive application. The large memory demand mainly 
arises due to storage of the elasticity  matrix and the tangent elasto-hydrodynamics linear system (\ref{eq:elastoHydrodynamic}), both of which has a size of the order of $(nx\times ny)^{2}$ elements,
where $nx$ and $ny$ are the total number of elements in the $x$ and $y$
directions of the grid respectively. For example, a simulation with $200$ cells
in both the $x$ and $y$ directions requires about $\sim27$GB of
storage. Keep in mind that due to the use of the HF tip asymptotic solutions, ILSA requires a much smaller number of elements to achieve
the same level of accuracy as compared to traditional numerical methods used in fracture mechanics. For example, for a radial fracture benchmark, ILSA requires 
about $\sim200$ times less elements as compared to a finite element
based method to achieve the same level of accuracy \cite{LePe13}. To reduce the memory requirement, PyFrac stores the elasticity matrix in single precision which brings down the memory requirement from $\sim27$GB to
$\sim20$GB in the case of a 200$\times$200 grid. 
Note that the pseudo-Toeplitz structure of the elasticity matrix for a rectangular grid could be used to further reduce the memory requirement but would require the development of specific matrix-matrix and matrix-vector dot products routines in order to efficiently built the elasto-hydrodynamic system.

\subsection{Classes}
Although PyFrac makes use of object orientated programming to structure  the code, we use it 
cautiously in order to avoid computational overhead.
Central to the code is the {\fontfamily{pcr}\selectfont Fracture} class which stores the state of the fracture at a given time. Advancing of the solution in time is done by a class
denoted as {\fontfamily{pcr}\selectfont Controller}. We briefly describe the different classes and their methods. 
\begin{itemize}
\item {\fontfamily{pcr}\selectfont Fracture}: This class stores the information about the state of the fracture at a given time including the width, the fluid pressure, the net pressure, the location of the front, the velocity of the front, the classification of the
grid cells (Channel, survey or Tip) containing the fracture and some other parameters. Methods to initialize a fracture to be advanced in time are also provided in the class. A fracture can be initialized with a footprint of arbitrary geometry with a given  pressure or with limiting case analytical solutions for a set of radial and height contained fracture geometries. Visualization methods  
for different fracture variables such as the footprint, width, pressure
and others are also available. 
\item {\fontfamily{pcr}\selectfont CartesianMesh}: This class defines a regular rectangular mesh with
the given dimensions. The class is fairly simple as the mesh is regular
and fixed. It stores the coordinates of the cell centers where the
fracture width and pressure are evaluated. The coordinates of the
vertices and their connectivity to the cells is also stored. A function
to visualize the mesh in 2D and 3D is provided by the class.
\item {\fontfamily{pcr}\selectfont Controller}: The Controller class is responsible for advancing the solution via appropriate time stepping according to the directives given 
by the {\fontfamily{pcr}\selectfont SimulationProperties} class. Re-attempts are made with slightly smaller or larger time steps in the case where a time step fails to converged in the prescribed number of iterations. If a time step fails even after re-attempts, the simulation is started again from the state of the fracture before the last five time steps. The Controller class is also responsible for saving the result to the file system for further post-processing or for visualizing the results during the simulation. 
\item {\fontfamily{pcr}\selectfont Property} classes: Property classes is a set of classes describing the material and fracturing fluid properties, and other simulation parameters.
PyFrac defines and use the following property classes:
\begin{itemize}
\item {\fontfamily{pcr}\selectfont MaterialProperties}: The parameters describing the properties of the material are stored in this properties class. These parameters include the plane strain modulus, the fracture toughness, the Carter's leak off coefficient, the in-situ confining stress, the grain size and the minimum residual width. The parameters that can vary spatially can be provided in the form of an array giving their value for each cell of the grid, or in the form of a function taking the coordinates as argument and returning the value of the parameter at the given coordinates. If the material has an anisotropic fracture toughness, the variation of the fracture toughness with
the propagation direction can be specified in the form of a function.
\item {\fontfamily{pcr}\selectfont FluidProperties}: This class stores the parameters describing the properties of the injected fluid such as the viscosity, compressibility and its density. A flag controls the use of friction factor model to account for the occurrence of turbulent flow.
\item {\fontfamily{pcr}\selectfont InjectionProperties}: This class stores the injection parameters such as the injection rate history and the source location. Variable injection
rate can be specified by giving a list of injection rate values and the time period for which they apply.
\item {\fontfamily{pcr}\selectfont SimulationProperties}: This class stores all the necessary directives for the controller to run the simulation. These include the numerical parameters such as the tolerances and maximum allowable iterations 
for the different iterative loops of  the algorithm, the parameters for simulation time and time stepping, the directives for output and visualization, the type of solvers to be used and some other miscellaneous directives. A total of about 45 simulation parameters and directives are stored by this class, details of which can be seen in the source code and its documentation.
\item {\fontfamily{pcr}\selectfont PerformanceProperties}: This class stores the performance data of an iteration that can be used to profile the computational performance of the code.
\item {\fontfamily{pcr}\selectfont PlotProperties}: This class stores the parameters to be used for plotting
the post-processed results. The purpose of the class is to bundle
the parameters which will be used for plotting across all of the visualization routines of PyFrac.
\end{itemize}
\end{itemize}

\subsection{Additional features}
\begin{itemize}
\item Post processing and visualization: PyFrac provides all the necessary 
routines to post process and visualize the results. Fracture parameters
including footprint, width, fluid and net pressure, front velocity, maximum/minimum/mean
distance between injection point and front, fluid velocity, fluid
flux, Reynold's number, fracture volume, leaked off volume, fracturing
efficiency and aspect ratio of the fracture can be visualized. These
parameters can be plotted on the complete mesh, on a slice perpendicular
to the plane containing the fracture, or on a specified point in the
spatial domain. The routines utilize the matplotlib library. Apart
from these routines to plot the numerical results, routines are also
provided to visualize  analytical solutions for radial and height contained fracture geometries.
\item Computational Performance profiling: PyFrac has the capability to
monitor and record the performance of computationally costly routines,
which can help in assessing the overall performance of the code. These
computational routines typically involves an iteration such as the
fracture front iteration, the fixed point iteration or a root finding
algorithm. The performance data is stored in PerformanceProperties
class objects, which are stored in the form of a tree. A node stores
performance data such as the CPU time taken, the number of iterations
taken to converge, a list of norms evaluated after each iteration
and some related information for a particular run of a routine. For
each node, the performance data for subroutines under that routine
are stored in the deeper branches of the tree. For example, for each
node storing information about a time step attempt, the performance
data for the fracture front iteration is stored as a branch in a deeper
level. For each of this fracture front node, the performance data
for the fixed point iteration is stored in a still deeper level of
the tree. Functions to post-process the saved performance data and
its visualization are also provided.
\item Remeshing: Once the fracture front reaches the end of the computational domain, PyFrac provides the capability to remesh it to automatically increase its size by a factor. This is done by making a new mesh with the same number of cells
as the original mesh but having dimensions scaled up by a given factor (2 by default).
By doing this, the elasticity matrix of the new scaled mesh can be
evaluated by dividing the old elasticity matrix with the scaling factor, allowing to avoid its revaluation upon remeshing. The variables are projected onto the new coarse mesh in a way to ensure proper volume conservation. The current fracture front is also projected onto the new mesh by interpolating the level set from the old mesh onto the new mesh and constructing the front on the new mesh using the Fast Marching Method. Remeshing allows to simulate fracture propagation over long time and lengthscale with a relatively small computational cost. Of course, such a feature must be used with care when accounting for the presence of material or in-situ stress heterogeneities.
\item symmetric fracture: The memory and computational requirements can
be significantly reduced for strictly symmetric fractures. For the case of an inviscid fracturing fluid (zero viscosity), PyFrac provides
the possibility of solving for only one quadrant for fractures that
are symmetric along the $x$ and $y$ axes. This reduce the memory requirement by a factor of $\sim16$ and computational
requirement by a factor of $\sim64$ for the linear system solver,
the most computationally costly subroutine of the code.
\end{itemize}

\section{Examples}
\subsection{Radial hydraulic fracture verification test}
\begin{figure}
\begin{centering}
\includegraphics[width=0.5\textwidth]{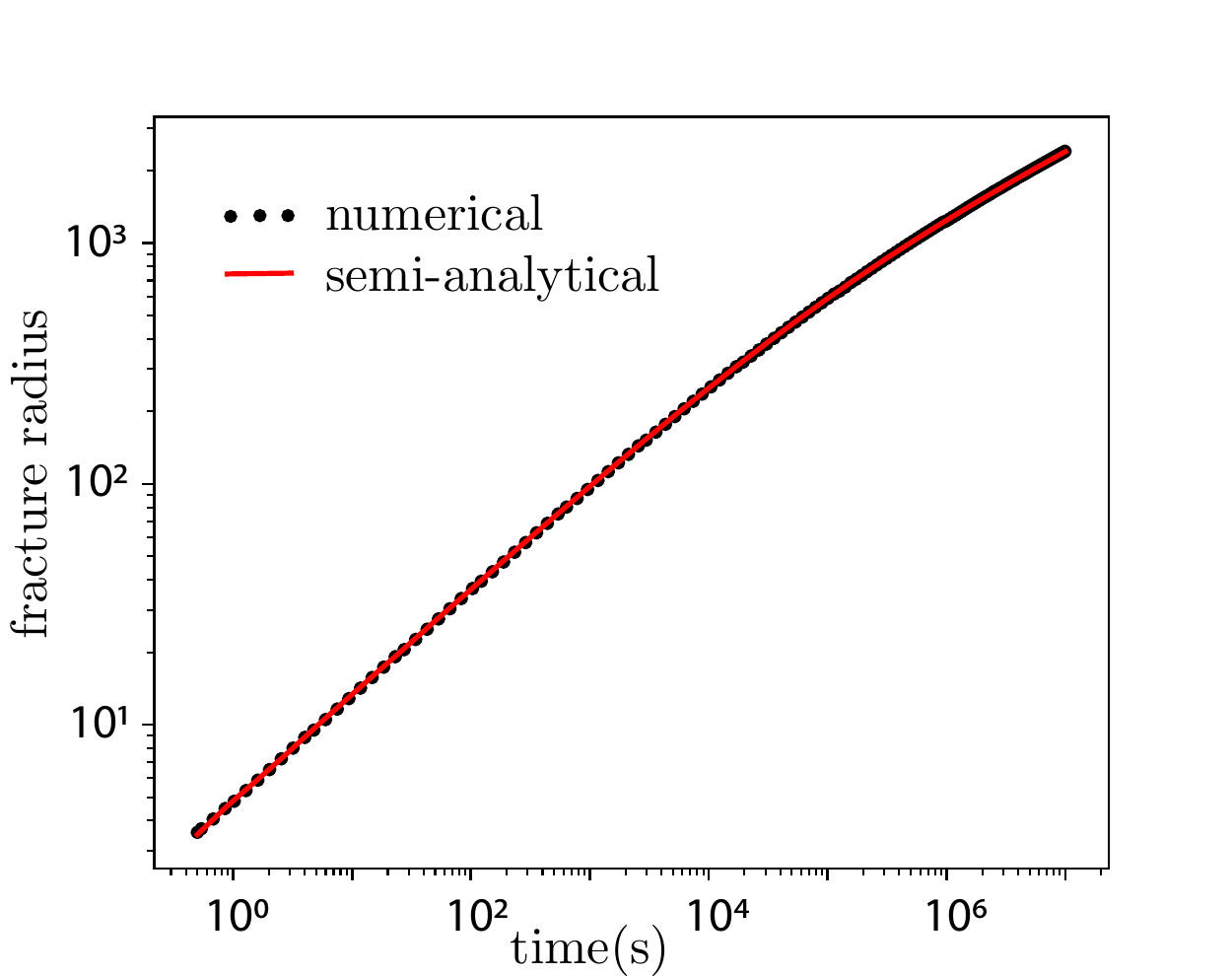}\includegraphics[width=0.5\textwidth]{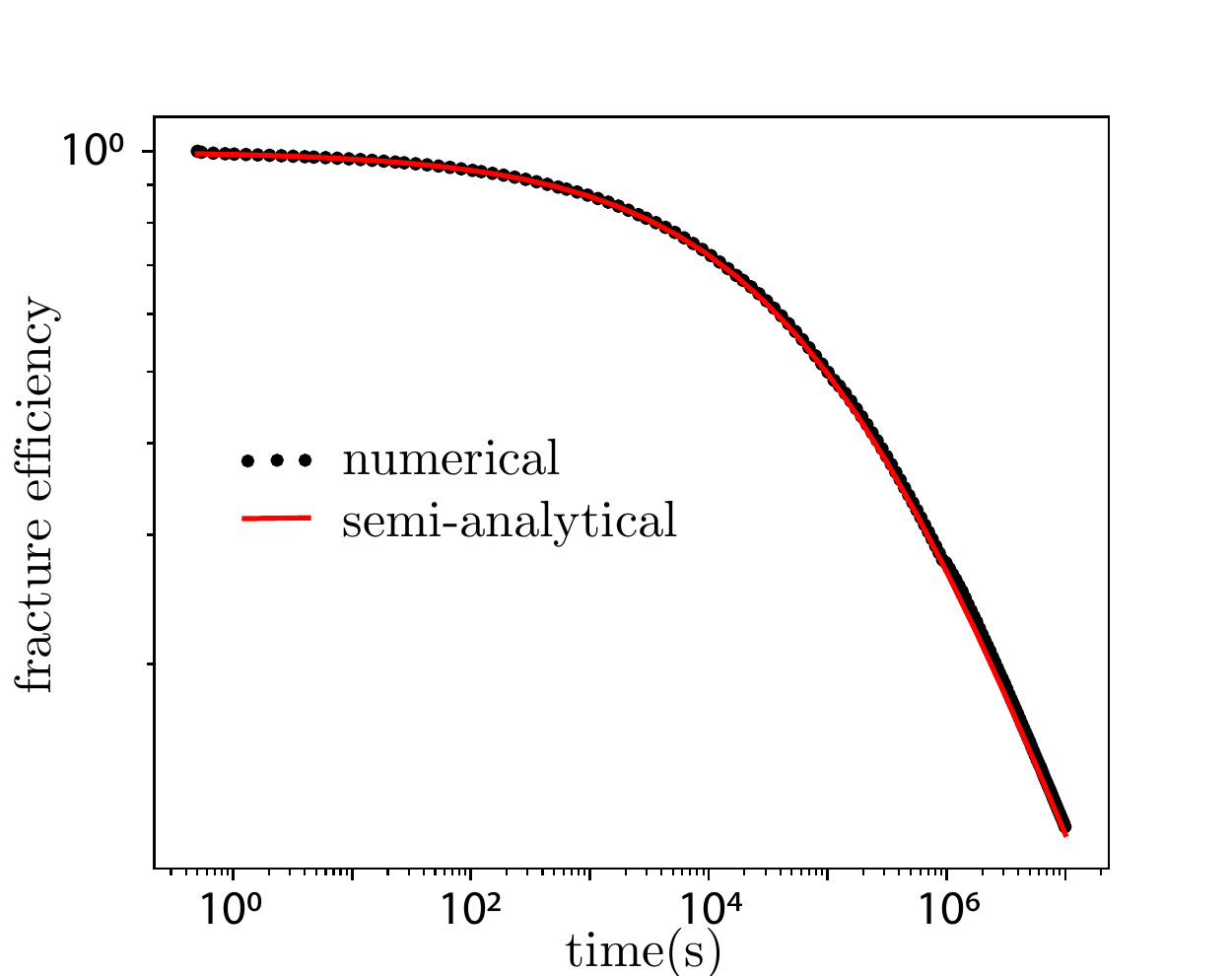}
\par\end{centering}
\caption{Viscosity storage to toughness leak off transition of a penny shaped hydraulic fracture. The fracture radius (left) and fracture efficiency (right) obtained using PyFrac are displayed against the semi-analytical solutions \cite{mady03,Dontsov2016} obtained with the code provided by \citet{dryad_gh469} for reference. \label{fig:radial_r_eff}}
\end{figure}

We first demonstrate the accuracy of PyFrac on the case of a radial (penny-shaped) hydraulic fracture propagating in a uniform permeable medium. The fracture  starts propagating in the viscosity dominated regime and gradually transitions to toughness and finally to leak-off dominated regime (see \citet{mady03,Dont16} for discussion of the reference solution). Here, a simulation is performed for a medium having a fracture toughness $K_{Ic}$ of $0.156$~MPa$\sqrt{\textrm{m}}$, a plane strain elastic modulus $E^\prime$ of $35.2$GPa and a leak-off coefficient $C_L$ of $0.5\times10^{-6}$~m/$\sqrt{\textrm{s}}$. The incompressible fluid ($c_f=0$) driving the fracture growth has a viscosity $\mu$  of $8.3\times10^{-5}$~Pa.s and is injected at a constant rate $Q_o$ of $ 0.01~\textrm{m}^{3}/\textrm{s}$. The simulation is started with a square domain of [$-5$, $5$, $-5$, $5$] meters divided into $41$ cells in both the $x$ and $y$ directions.

\begin{figure}
\begin{centering}
\includegraphics[width=0.5\textwidth]{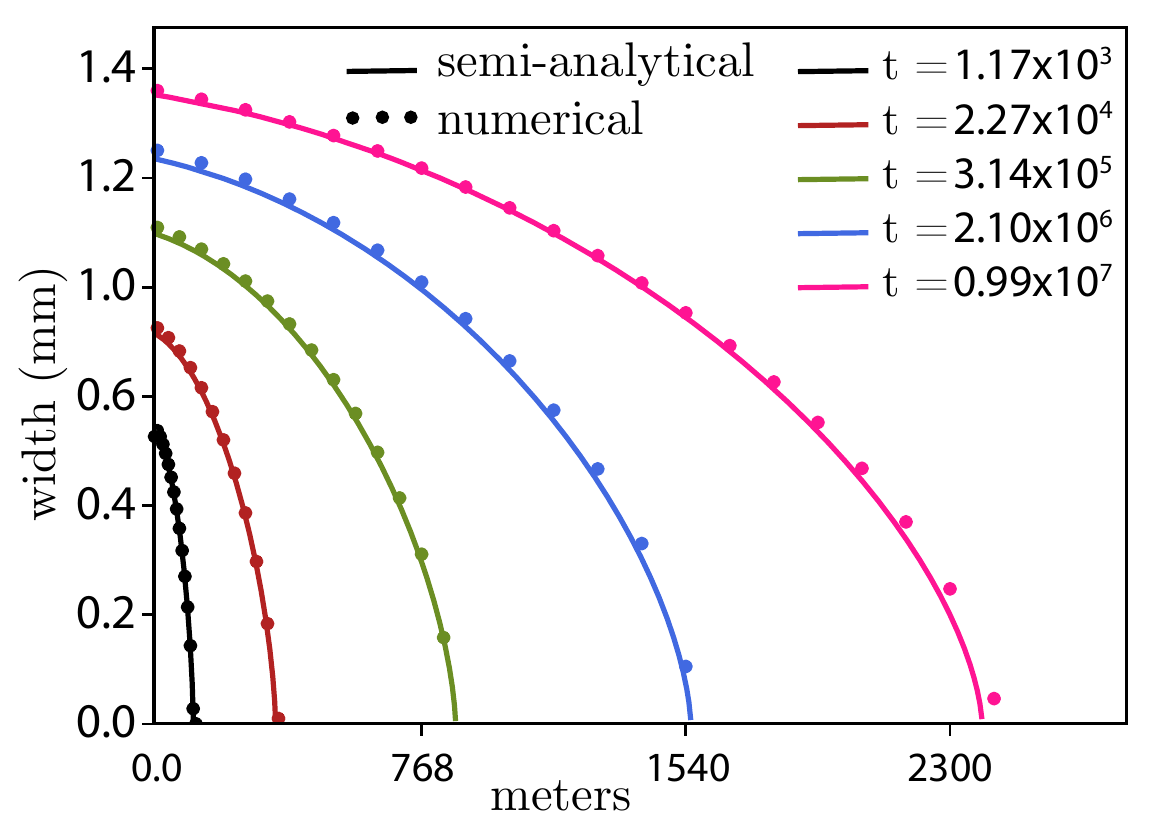}\includegraphics[width=0.5\textwidth]{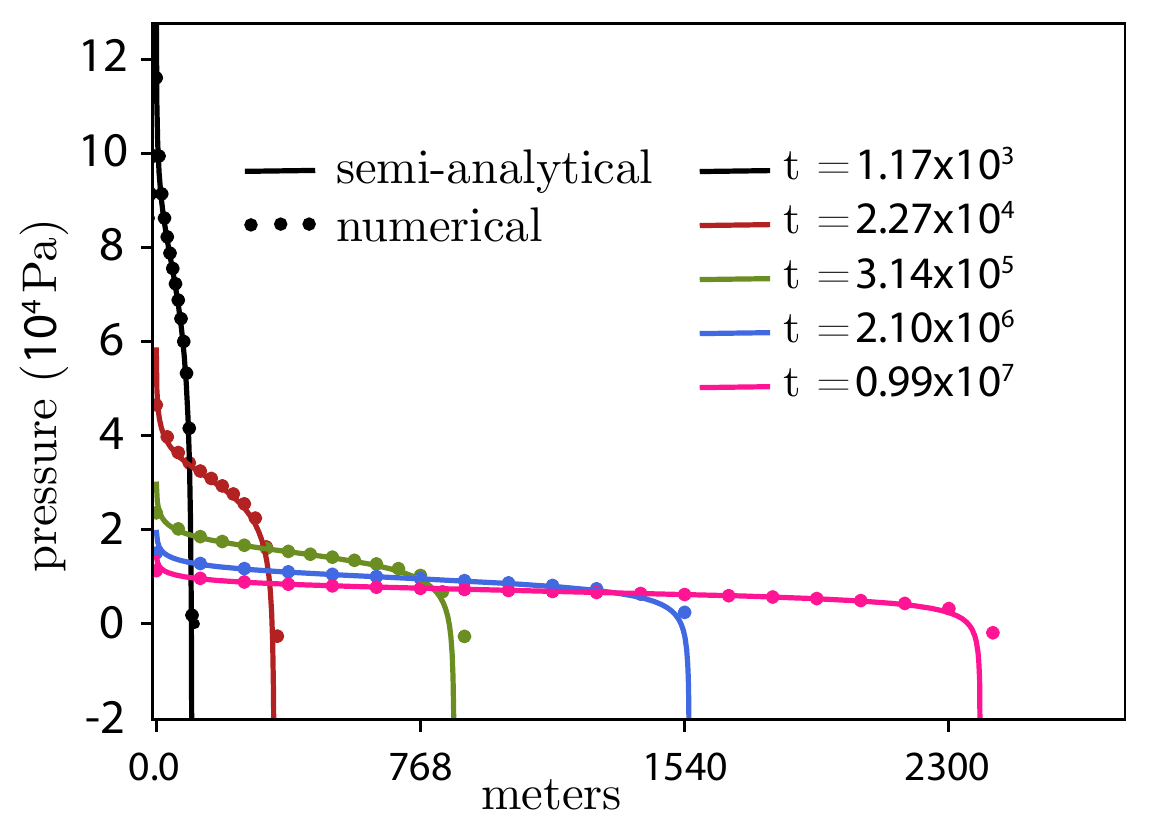}
\par\end{centering}
\caption{The width (left) and pressure (right) profiles at selected times ($t = [1170$, $2270$, $313775$, $2096374$, $9929186]$ seconds) along the positive x-axis. The reference solution \cite{mady03} obtained with the code provided by \citet{dryad_gh469} at these times are also shown for reference. \label{fig:radial_p_w}}
\end{figure}

Figure \ref{fig:radial_r_eff} displays the evolution of fracture radius (left) and fracture efficiency (right) with time. The fracture efficiency is defined as the ratio of the volume of the fluid currently present in the fracture to the total volume injected. Figure \ref{fig:radial_p_w} displays the width (left) and pressure (right) profiles along slices made at the positive x-axis  at $t = [1170$, $2270$, $313775$, $2096374$, $9929186]$ seconds. A very good agreement between the numerical solution and the reference solution can be seen in both figures.

\subsection{Height contained hydraulic fracture}

\begin{figure}
\begin{centering}
\includegraphics[width=0.5\textwidth]{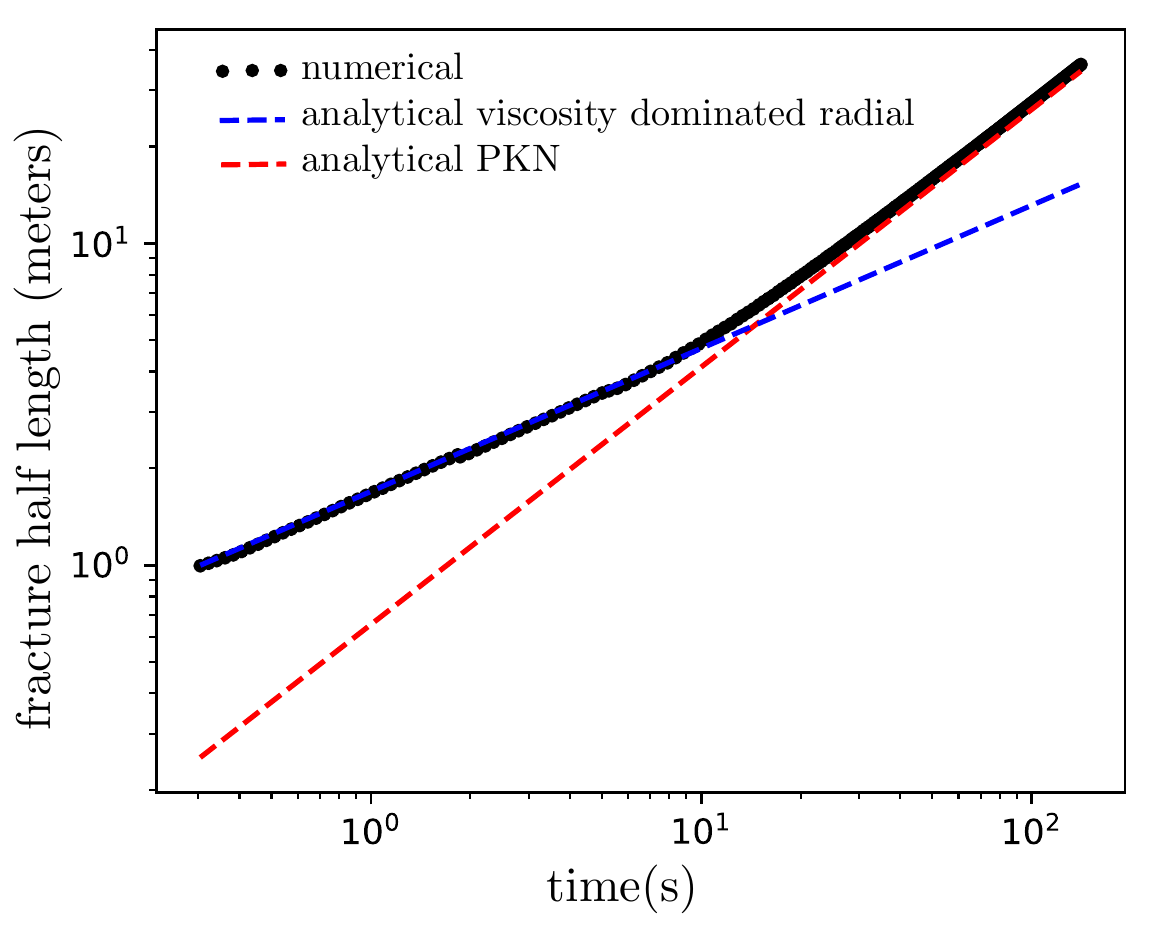}\includegraphics[width=0.5\textwidth]{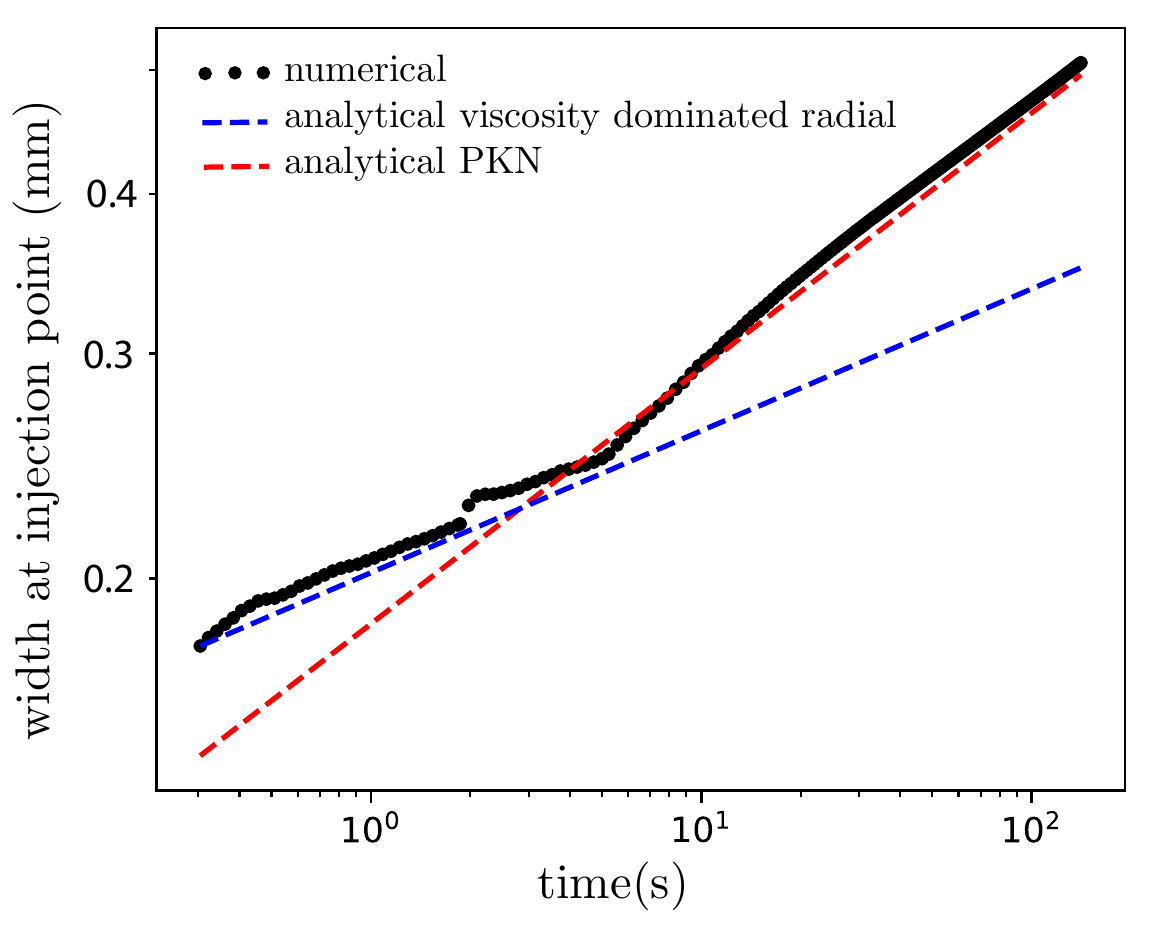}
\par\end{centering}
\caption{Transition from radial to PKN (height contained) geometry. The time evolution of the fracture half length along x-axis (left) and the fracture width at injection point (right) calculated with PyFrac. The analytical viscosity dominated radial and PKN solution are also shown for reference. \label{fig:PKN_r_w}}
\end{figure}
 
\begin{figure}
   \makebox[\textwidth]{
     \includegraphics[width=\textwidth]{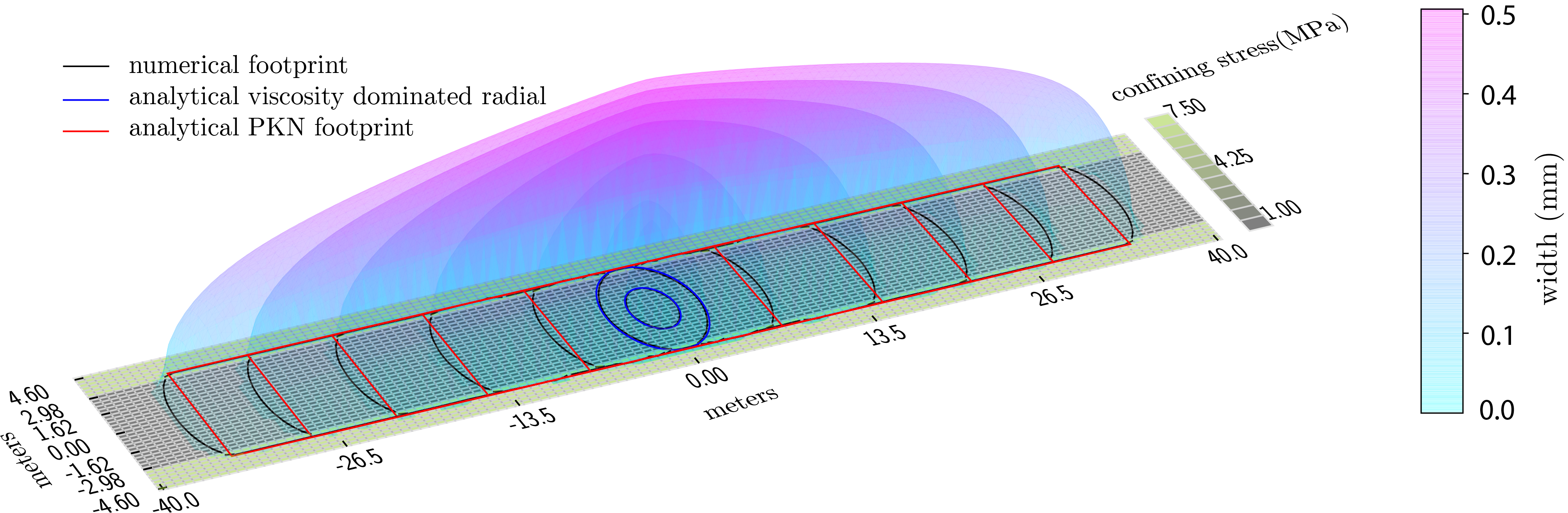}
   }
   \caption{Fracture footprint and the fracture width at $t=[1$, $5$, $20$, $50$, $80$, $110$, $140]$ seconds for the height contained fracture propagation example. The solution initially agrees with the viscosity dominated radial solution (shown in blue) and later on transitions to the PKN solution (shown in red).} \label{fig:PKN_footprint}
\end{figure}

This example simulates a hydraulic fracture propagating in a layer bounded with high stress layers from top and bottom, causing its height to be restricted to the height of the middle layer. The top and bottom layers have a confining stress of $7.5$Mpa, while the middle layer has a confining stress of $1$MPa (see figure \ref{fig:PKN_footprint}). The fracture initially propagates as a radial fracture in the middle layer until it hits the high stress layers on the top and bottom. From then onwards, it propagates with the fixed height of the middle layer.

The parameters used in the simulation are as follows:
\[
E^{\prime}=35.2\textrm{GPa},\;K_{Ic}=0,\;\mu=1.1\times10^{-3}\textrm{Pa.s},\;Q=0.001\textrm{m}^{3}/\textrm{s}.
\]
A rectangular domain with dimensions of $[-20, 20, -2.3, 2.3]$ meters is used for initial propagation. As the fracture grows and reach the end of the domain, a remeshing is done to double the size of the domain to $[-40, 40, -4.6, 4.6]$. The domain is divided into $125$ cells in the $x$ direction and $35$ cells in the $y$ direction.

Figure \ref{fig:PKN_r_w} shows the evolution of the fracture length (left) and fracture width at injection point (right) with time. Expectedly, the solution first follows the viscosity dominated radial fracture solution and then transitions to height contained regime for which the classical PKN \cite{PKN61} solution is applicable. The error introduced in the solution at about $2$ seconds is due to remeshing. Figure \ref{fig:PKN_footprint} shows the footprint and the width of the fracture at $t=[1$, $5$, $20$, $50$, $80$, $110$, $140]$ seconds. It can be seen that the footprint matches closely to the radial fracture solution initially and then to the PKN solution.

\subsection{Lateral spreading of a Dyke at neutral buoyancy}

\begin{figure}
\begin{centering}
\includegraphics[width=\textwidth]{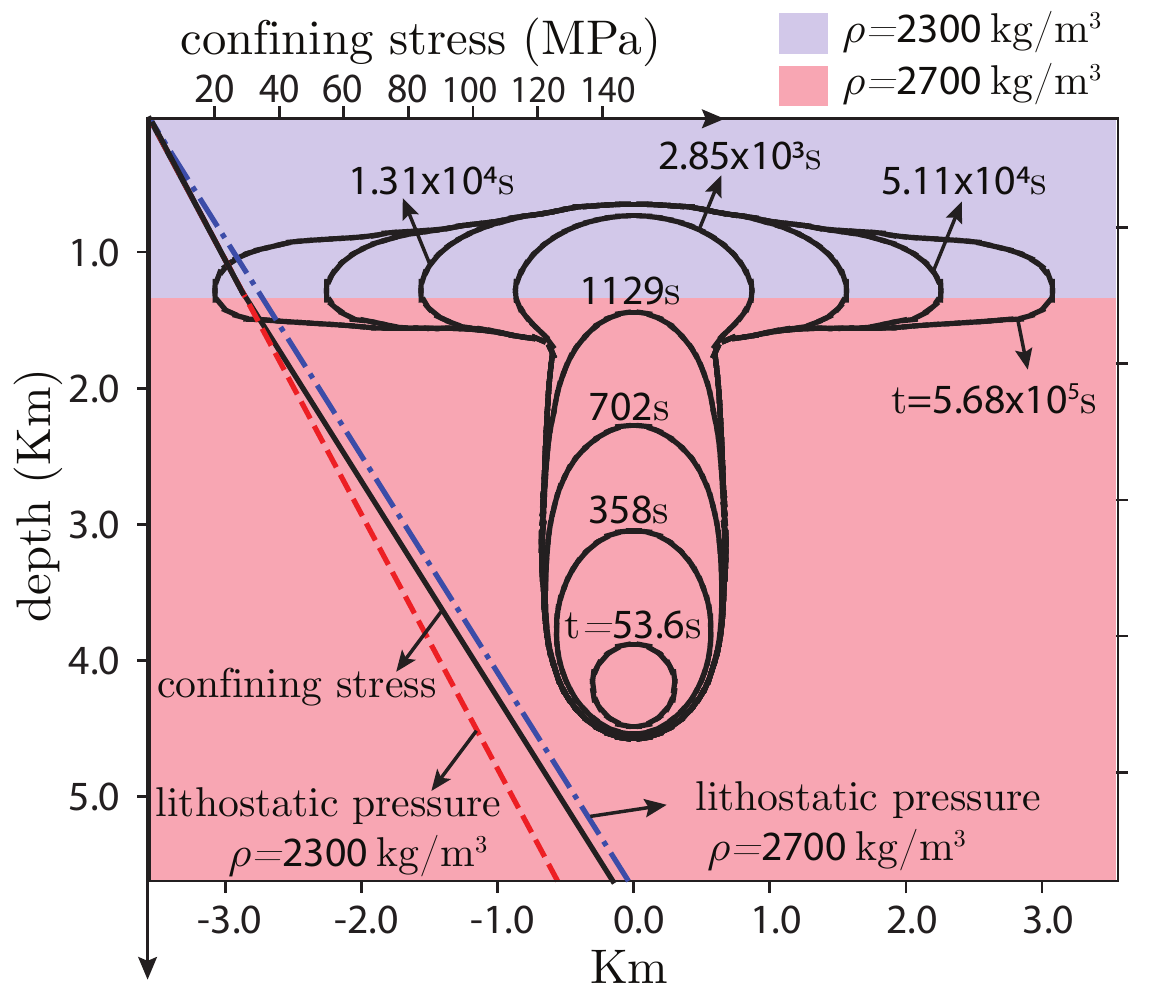}
\par\end{centering}
\caption{The footprint of the dyke at $t=[53.59,$ $357.61,$ $702.45,$ $1129.1,$ $2855.14,$ $13173.53,$ $51145.96,$ $568317.8]$ seconds. The two layers with different densities and the resulting in-situ confining stress (black line) as a function of the depth is also shown. The dotted red and blue lines show the lithostatic pressure for the case of homogeneous rocks with densities of $2300\textrm{Kg/m}^{3}$ and $2700\textrm{Kg/m}^{3}$ respectively. \label{fig:LS_footprint}}
\end{figure}

This example demonstrates the capability of PyFrac to simulate buoyancy driven fractures. Here, we simulate propagation of a dyke after a pulse injection of basaltic magma at a depth of $4.2$Km. The magma fractures the surrounding rock towards the surface as a dyke and  reaches a layer with lower density at a depth of $1.3$Km: actually reaching neutral buoyancy. As a result, the propagation is then arrested vertically and the dyke spreads horizontally. For this simulation, we take values of the rock and magma parameters similar to the one reported in \cite{TrPi10}. 
We notably set the plane strain modulus of the rock to $E^{\prime}=1.2\textrm{GPa}$ and its fracture toughness to  $K_{Ic}=6.5\textrm{Mpa}\sqrt{\textrm{m}}$. The density of the rock is taken as $\rho_r=2700\textrm{Kg/m}^{3}$ for the lower layer (below $1.3$km from the surface) and  $\rho_r=2300\textrm{Kg/m}^{3}$ for the upper layer (see figure \ref{fig:LS_footprint}). The pulse release of magma is done by injecting with an injection rate of $2000\textrm{m}^{3}/\textrm{s}$ for the first $500$ seconds amounting to a total injected volume of $10^6\textrm{m}^{3}$. For magma, we have used a density of  $\rho_f=2400\textrm{Kg/m}^{3}$ and a viscosity of $30$Pa.s. The simulation is performed with a mesh having $83$ cells in both $x$ and $y$ dimensions.

\begin{figure}
\begin{centering}
\includegraphics[scale=0.6]{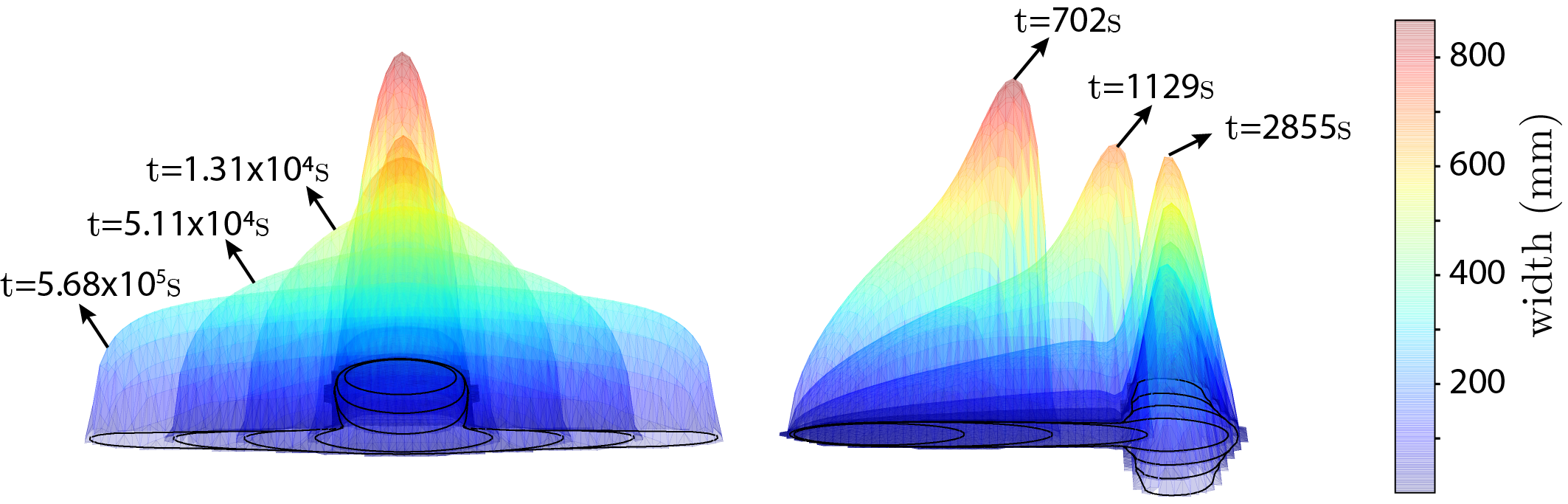}
\caption{The opening width of the dyke after the pulse injection at $t=[702.45$, $1129.1$, $2855.14$, $13173.53$, $51145.96$, $568317.8]$ seconds. \label{fig:LS_width}}
\par\end{centering}
\end{figure}

Figure \ref{fig:LS_footprint} shows the evolution of the footprint of the dyke as it propagate with time (for $t=[53.59,$ $357.61,$ $702.45,$ $1129.1,$ $2855.14,$ $13173.53,$ $51145.96,$ $568317.8]$ seconds). The confining stress vs depth along with the lithostatic pressure profile (taken as the minimum in-situ stress in this example) induced by both the low and high rock densities is also shown for reference.
It can be seen that the dyke initially propagates upwards rapidly ($v\sim 2.7$m/s) but its velocity drops with time. $14$ hours after release, as it is spreading laterally, it has almost stopped ($v\sim 1$cm/s). It comes to a complete arrest at around $157$ hours after the pulse injection. The evolution of the width of the dyke after the pulse release is shown in figure \ref{fig:LS_width}.


\subsection{Fracture closure}

\begin{figure}
\includegraphics[width=\textwidth]{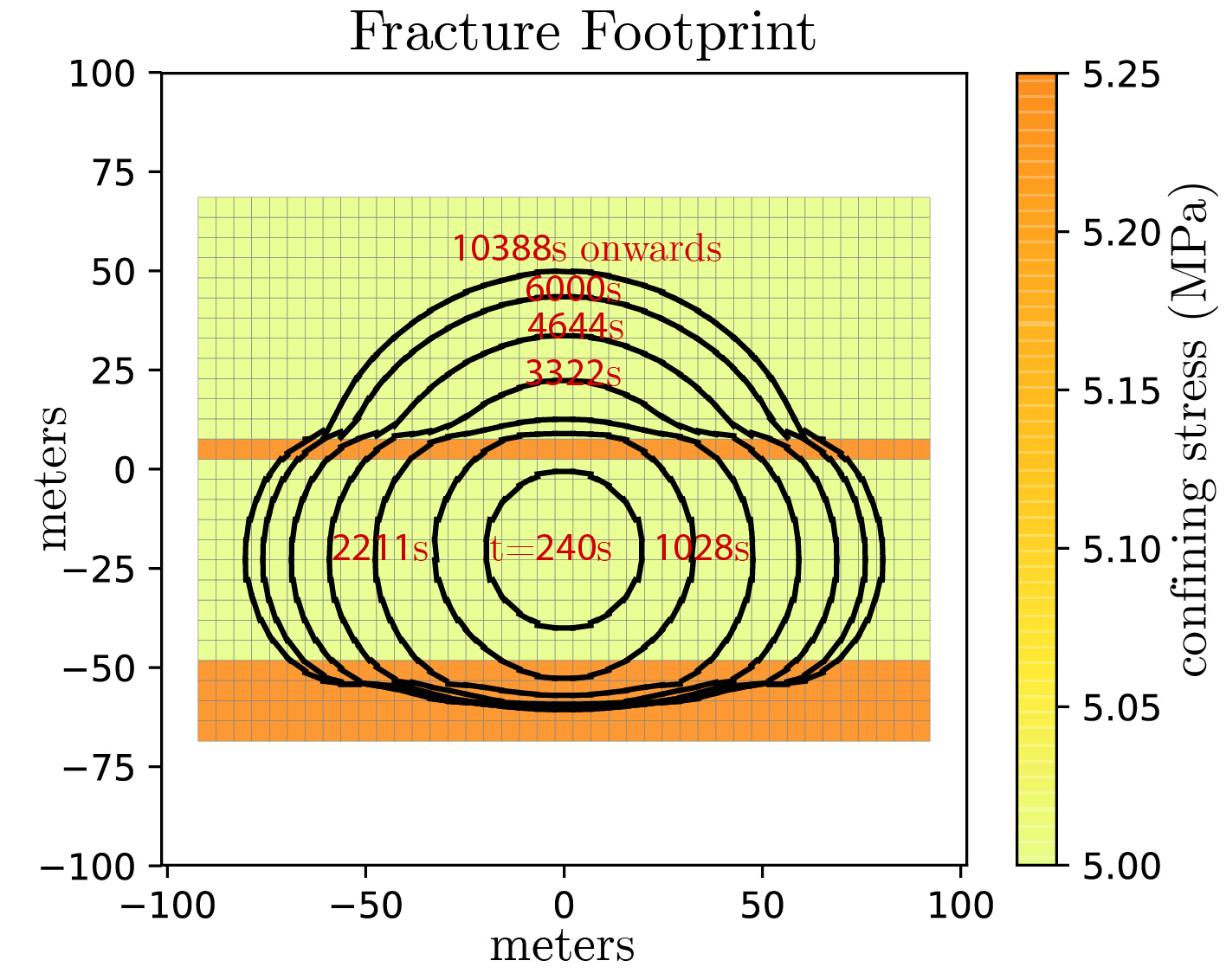}
\centering{}
\caption{The footprint of the fracture at selected times for the fracture closure example, including the time at which the injection stops ($6000$s). The fracture stops propagating after $10388$s.  \label{fig:FC_footprint}}
\end{figure}

In this example, we show the capability of PyFrac to handle fracture closure. The simulation consists of a 100 minutes injection of water at the rate of $10^{-3}\textrm{m}^3/\textrm{s}$ into a rock with a plane strain elastic modulus of $E^{\prime}=42.67\textrm{GPa}$ and fracture toughness of $K_{Ic}=0.5\textrm{Mpa}\sqrt{\textrm{m}}$. The minimum aperture $w_r$ upon closure is set to 1 micron.
The fracture is initiated in a layer that is bounded by layers having higher confining stress. The layer on top is set to have a small height, allowing the fracture to break through and accelerate upwards in another layer (see figure \ref{fig:FC_footprint}). The rock is taken to be permeable with a Carter's leak off coefficient of $C_L=10^{-6}\textrm{m}/\sqrt{\textrm{s}}$. The simulation is performed in a rectangular domain with the dimensions of $[-90$, $90$, $-66$, $66]$ meters, which is divided into $41$ and $27$ cells in the $x$ and $y$ directions respectively. 

\begin{figure}
\includegraphics[width=0.55\textwidth]{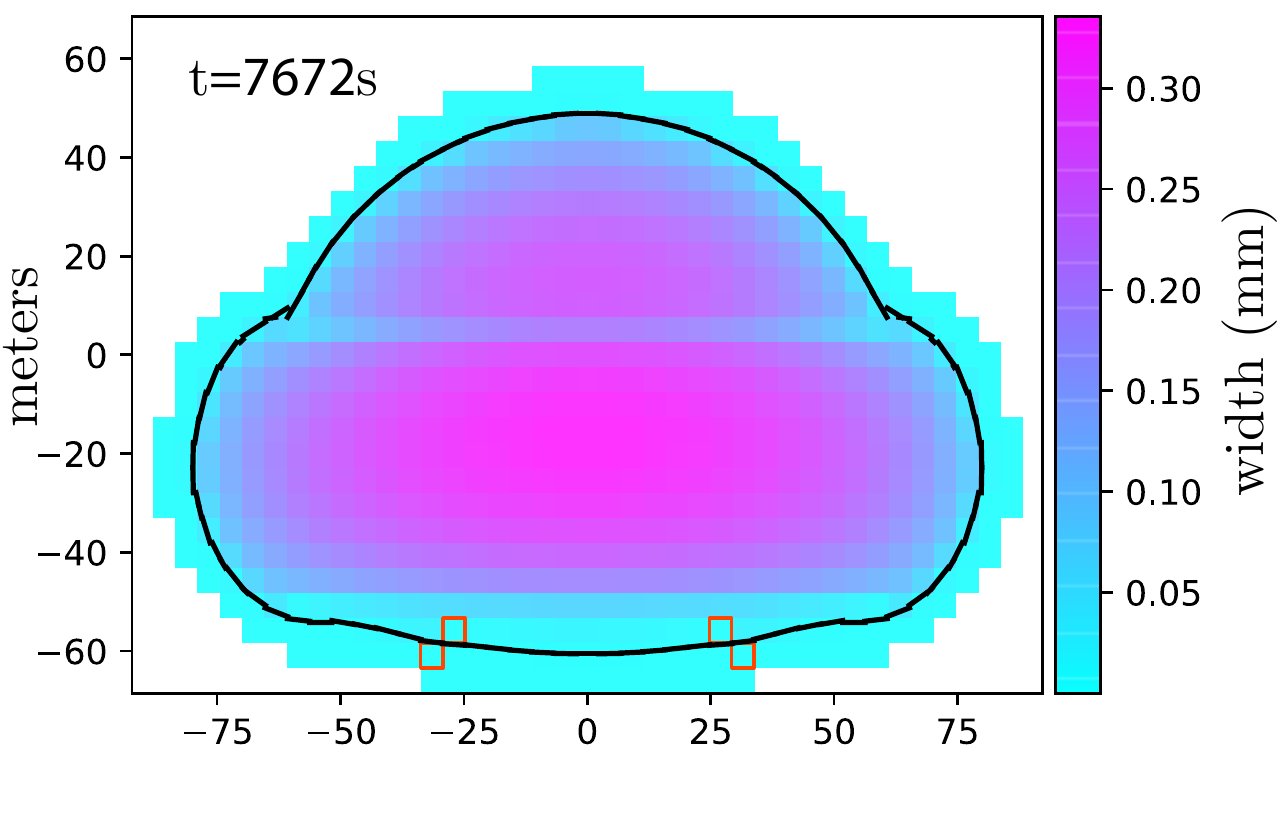}\includegraphics[width=0.55\textwidth]{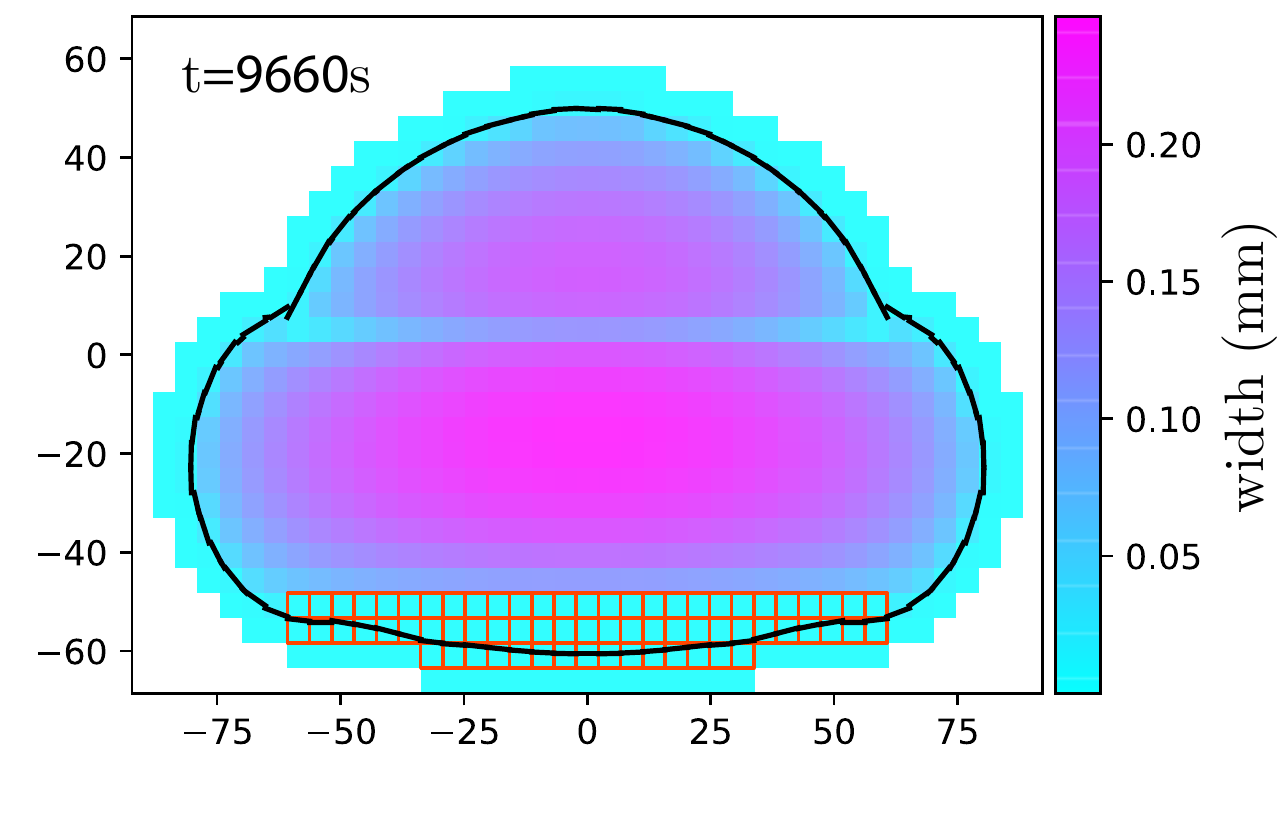}

\includegraphics[width=0.55\textwidth]{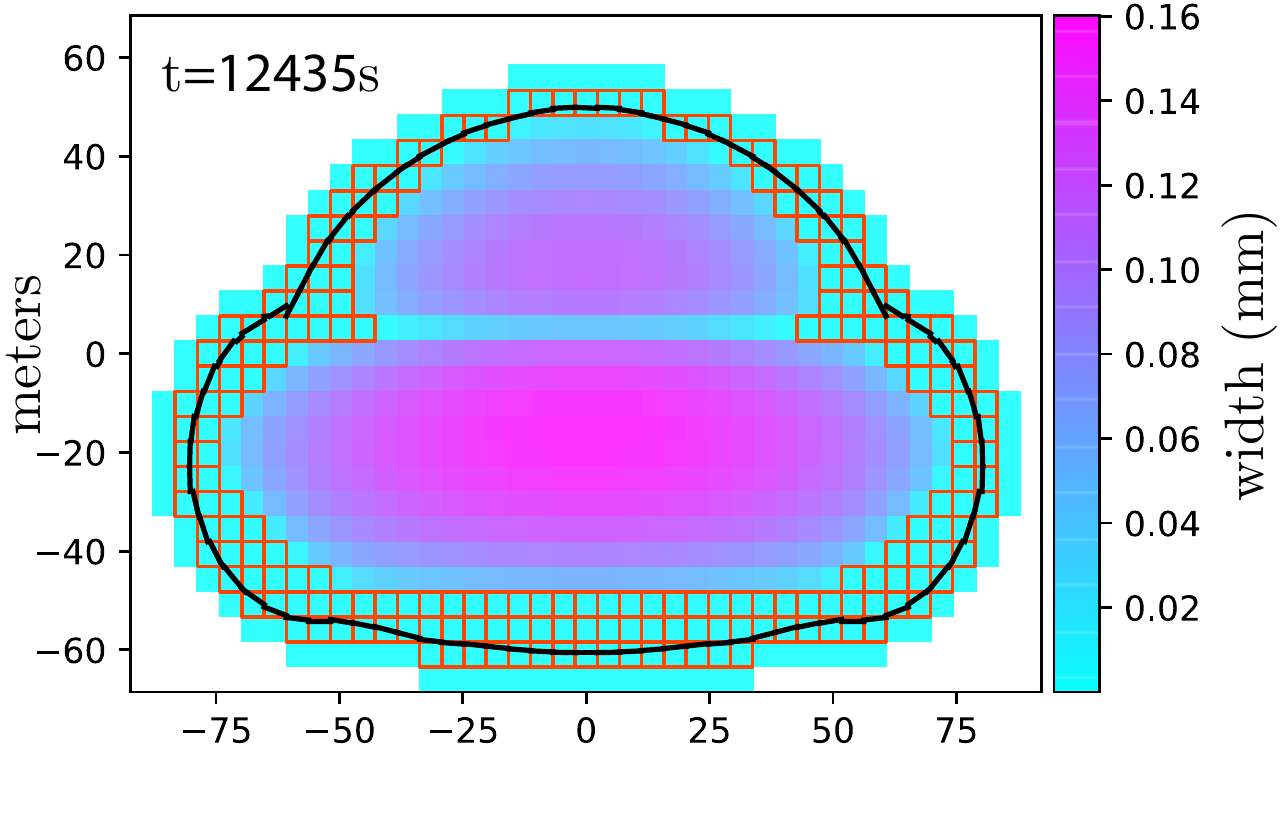}\includegraphics[width=0.55\textwidth]{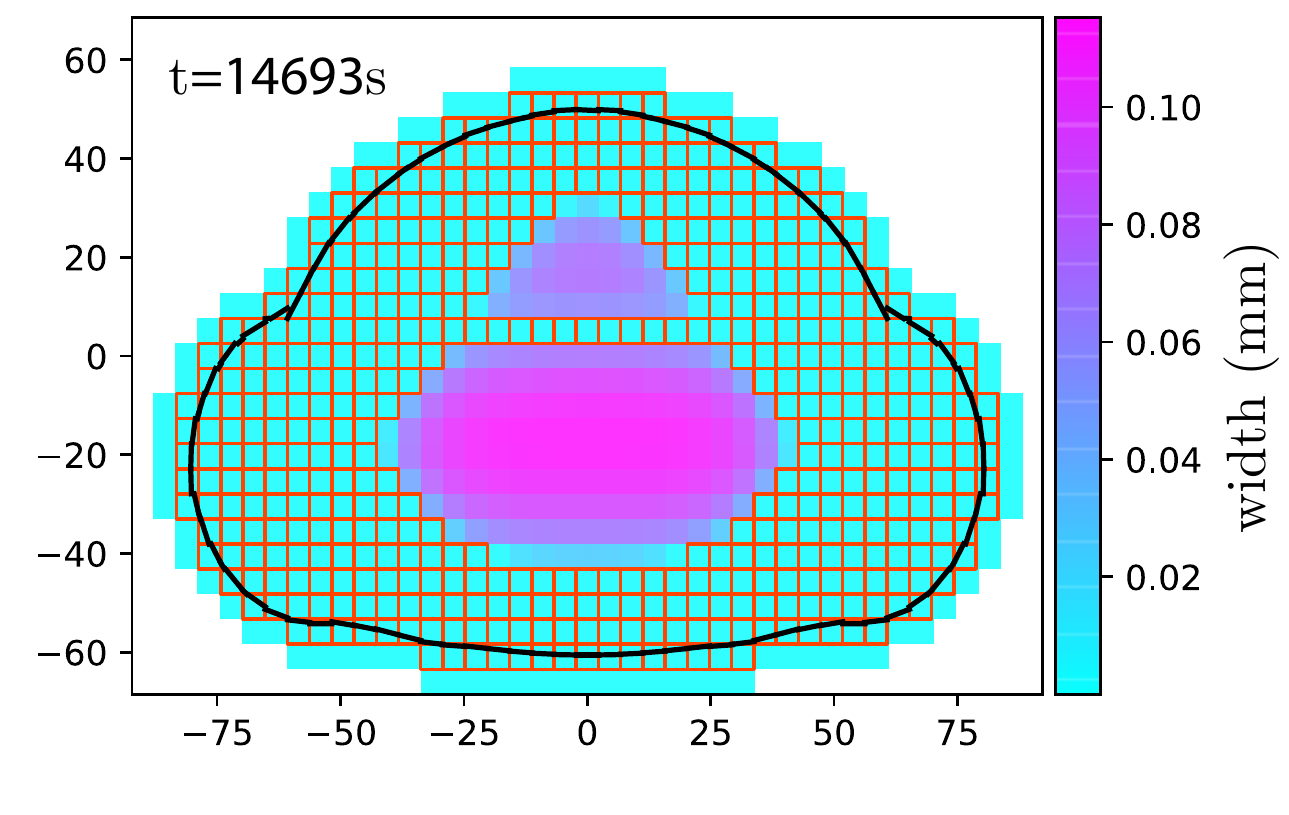}

\includegraphics[width=0.55\textwidth]{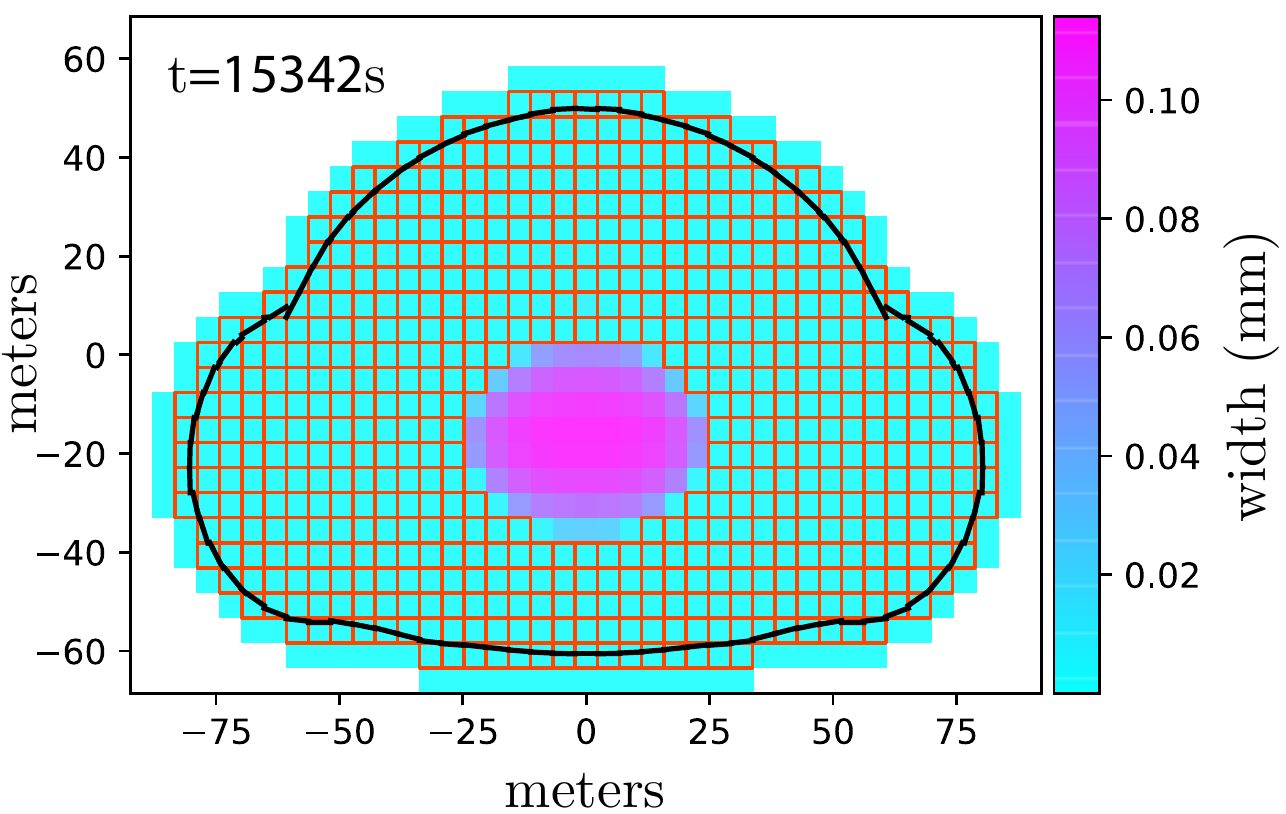}\includegraphics[width=0.55\textwidth]{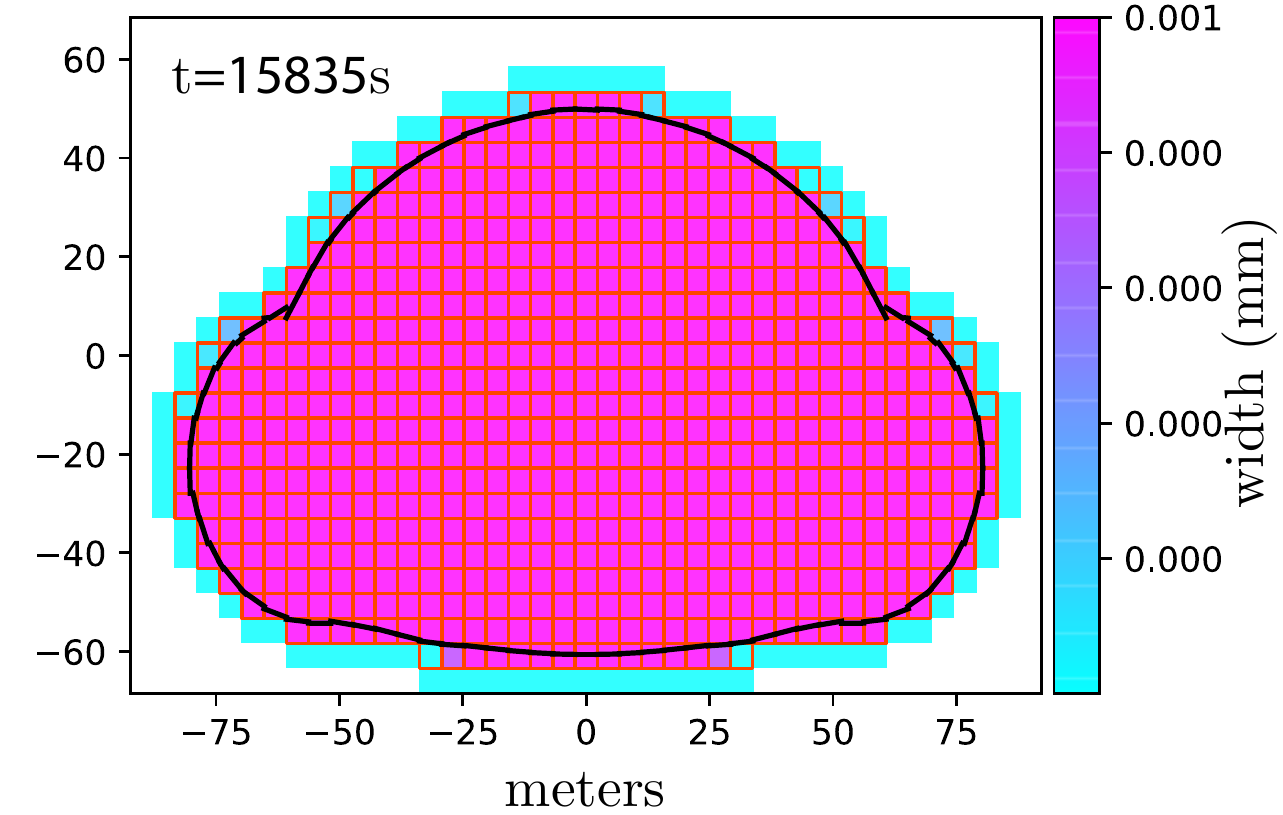}

\caption{Fracture width at different times for the fracture closure example. The cells outlined in red are mechanically closed ($w_r=10^{-3}$ mm here). The fracture starts to close at around $t=7672$s in the bottom high stress layer which gets fully closed at $t=9660$s. Closure then continues from the tip inwards ($t=12435$s) until both the high stress layers are closed at $t=14693$s, dividing the fracture into two open regions. The complete fracture finally fully closes at $t=15835$s. \label{fig:closure}}
\end{figure}

Figure \ref{fig:FC_footprint} shows the footprint of the fracture at $t$=[$240$, $1028$, $2211$, $3322$, $4644$, $6000$, $10388$] seconds in combination with the corresponding in-situ confining stress. It can be seen that the fracture continues to slowly grow even after the injection has stopped at $6000$s until it comes to a complete stop at $10388$s. Due to fluid leak off, the fracture starts to close with time starting from $7672$s. Figure \ref{fig:closure} displays the  closure of the fracture with time. The cells displayed in red are mechanically closed 
(active width constraint). It can be observed that the fracture starts to close from its tip, first in the high stress layers. At around $14693$s, the fracture is fully closed in both the high stress layers, while two separate parts of the fracture are still open. As the fluid continue to leak off in the surrounding rock, the fracture finally fully closes at around $4.4$ hours ($t=15835$ seconds).

\section{Conclusions}

In this paper, we have presented PyFac, a python based implementation of
the implicit level set algorithm for the simulation of the growth of  a planar three-dimensional hydraulic fracture.
 The solver has been extensively verified against known semi-analytical solutions of planar HF growth in simple geometries (radial, height contained hydraulic fractures in different propagation regimes). Besides the examples described previously (whose scripts are included with the source code), a number of other tests and examples are also provided in this release. 
 PyFrac has a number of additional features compared to the original ILSA algorithm: in particular, the ability of model anisotropy in fracture toughness as well as elasticity, turbulent flow, heterogeneities of toughness among others. 
The current solver could also be extended to account notably for non-Newtonian fracturing fluid rheologies, multiphase fluids (e.g. proppant, slurry) and piece-wise heterogeneous elasticity. Additional code optimization could certainly further bring down the simulation cost. 
We hope PyFrac will foster benchmarking and reproducibility as well as the use of open-source codes in the hydraulic fracturing community.

\paragraph*{Acknowledgments}

This work was funded by the Swiss National Science Foundation under grant \#160577.

\paragraph*{CRediT Authors contributions\\}
{Haseeb Zia: Conceptualization, Methodology, Investigation, Software, Validation, Visualization, Writing -- original draft.
\\
Brice Lecampion: Conceptualization, Methodology, Software, Validation, Funding acquisition, Writing -- review \& editing.
}




\section*{REFERENCES}
\bibliographystyle{elsarticle-num}

\end{document}